\documentclass[twocolumn]{aastex62} 
\usepackage{hyperref,enumerate}
\input{colordvi}
\listofchanges
\hypersetup{
    unicode=false,                  % non-Latin characters in Acrobat book
    pdftoolbar=true,                % show Acrobat toolbar?
    pdfmenubar=true,                % show Acrobat menu?
    pdffitwindow=true,              % window fit to page when opened
    pdfstartview={FitH},            % fits the width of the page t
    pdftitle={R443130},             % title
    pdfauthor={Mardini},            % author
    pdfsubject={Astronomy},         % subject of the document
    pdfcreator={dvipdf},            % creator of the document
    pdfproducer={dvipdf},           % producer of the document
    pdfkeywords={metal-poor stars}, % list of keywords
    pdfnewwindow=true,              % links in new window
    colorlinks=true,                % false: boxed links; true: colored
    linkcolor=red,                  % color of internal links
    citecolor=blue,                 % color of links to bibliography
    filecolor=magenta,              % color of file links
    urlcolor=cyan,                  % color of external links
    breaklinks=true,
    linktocpage
}

\newcommand{\Teff}{$T_{\rm eff}$}  % Command for Teff, text mode
\newcommand{\kms}{km\,s$^{-1}$}

\newcommand{\Gyr}{\ensuremath{\mathrm{Gyr}}}

\shorttitle{Discovery of Five Bright Carbon-Enhanced Metal-Poor Stars}
\shortauthors{Mardini et al.}
\begin{document}

\title{METAL-POOR STARS OBSERVED WITH THE AUTOMATED PlANET FINDER TELESCOPE. I. DISCOVERY OF FIVE CARBON-ENHANCED METAL-POOR STARS FROM LAMOST}

\author[0000-0001-9178-3992]{Mohammad K.\ Mardini}
\affiliation{Key Lab of Optical Astronomy, National Astronomical Observatories, Chinese Academy of Sciences, A20 Datun Road, Chaoyang, Beijing 100102, China}
\affiliation{School of Astronomy and Space Science, University of Chinese Academy of Sciences, No.19(A) Yuquan Road, Shijingshan District, Beijing, 100049, China}
\author[0000-0002-0389-9264]{Haining Li}
\affiliation{Key Lab of Optical Astronomy, National Astronomical Observatories, Chinese Academy of Sciences, A20 Datun Road, Chaoyang, Beijing 100102, China}
\author[0000-0003-4479-1265]{Vinicius M.\ Placco}
\affiliation{Department of Physics, University of Notre Dame, Notre Dame, IN 46556, USA}
\affiliation{JINA Center for the Evolution of the Elements, USA}	
\author[0000-0002-8709-4665]{Sofya Alexeeva}
\affiliation{Shandong Provincial Key Laboratory of Optical Astronomy and Solar-Terrestrial Environment, Institute of Space Sciences, Shandong University, Weihai 264209, China }
\author{Daniela Carollo}
\affiliation{INAF Astrophysical Observatory of Turin - Italy}
\author{Ali Taani}
\affiliation{Physics Department, Faculty of Science, Al-Balqa Applied University,Jordan}
\author{Iminhaji Ablimit}
\affiliation{Department of Astronomy, Kyoto University, Kitashirakawa-Oiwake-cho, Sakyo-ku, Kyoto 606-8502, Japan}
\author{Liang WANG}
\affiliation{Max-Planck-Institut fur extraterrestrische Physik, Giessenbachstrasse, D-85748 Garching, Germany}
\author[0000-0002-8980-945X]{ Gang Zhao}
\affiliation{Key Lab of Optical Astronomy, National Astronomical Observatories, Chinese Academy of Sciences, A20 Datun Road, Chaoyang, Beijing 100102, China}
\affiliation{School of Astronomy and Space Science, University of Chinese Academy of Sciences, No.19(A) Yuquan Road, Shijingshan District, Beijing, 100049, China}	

\correspondingauthor{\\Haining Li [lhn@nao.cas.cn]\\Gang Zhao [gzhao@nao.cas.cn]}

\nocollaboration

\begin{abstract}
We report on the discovery of {five} carbon-enhanced metal-poor (CEMP) stars 
in the metallicity range of $-3.3<$  [Fe/H] $<-2.4$. These stars were selected from 
the LAMOST DR3 low-resolution (R$ \sim$ 2,000) spectroscopic database as 
metal-poor candidates and followed-up with high-resolution spectroscopy 
(R$ \sim$110,000) with the LICK/APF. Stellar parameters and individual abundances 
for 25 chemical elements (from Li to Eu) are presented for the first time.
These stars exhibit chemical abundance patterns that are similar 
to those reported in other literature studies of very and extremely metal-poor 
stars. One of our targets, J2114$-$0616, shows high enhancement in carbon 
([C/Fe]=1.37), nitrogen ([N/Fe]= 1.88), barium ([Ba/Fe]=1.00), and europium 
([Eu/Fe]=0.84). Such chemical abundance pattern suggests that J2114$-$0616 
can be classified as CEMP-r/s star. In addition, the star J1054+0528 can be 
classified as a CEMP-rI star, with [Eu/Fe]=0.44 and [Ba/Fe]=$-$0.52. The other stars in our 
sample show no enhancements in neutron-capture elements and can be classified as CEMP-no stars. 
We also performed a kinematic and dynamical analysis of the sample stars based on Gaia DR2 data. 
The kinematic parameters, orbits, and binding energy of these stars, show that J2114$-$0616 {is member}
of the outer halo population, while the remaining stars belong to the inner halo population but with an accreted origin. 
Collectively, these results add important constraints on the origin and evolution of CEMP stars as well as on their possible formation scenarios.

\end{abstract}


\section{Introduction} \label{sec:intro}

 Metal-poor stars play a crucial role in galactic archaeology, since they represent a fossil record of 
 the nucleosynthesis products of their progenitors 
 \citep{1995AJ....109.2757M, 2005ARA&A..43..531B,2017MmSAI..88..856C, 2017ApJ...848...85J}. 
 These stars contain detailed information about the past of their host systems, which can be used to 
 study the early Universe and the beginning of star and galaxy formation \citep{2015ARA&A..53..631F}.
 One of the primary tools to study metal-poor stars is the determination of their chemical abundances,
 elements heavier than lithium reflect the extent of chemical enrichment within its natal cloud 
 \citep{2002Natur.419..904C, 2005Natur.434..871F, 2015ARA&A..53..631F}.

 Many studies have indicated that metal-poor stars may show high carbon-to-iron ratios, and thus classified 
 as carbon-enhanced metal-poor stars \citep{2005ARA&A..43..531B, 2007ApJ...655..492A, 2016A&A...588A..37H, 2016ApJ...829L..24P, 2017MNRAS.471..404K,
 2017MNRAS.472..350R, 2018MNRAS.475.4781C, 2018A&A...614A..68C}. {Furthermore, CEMP stars can be divided into four sub-classes, according to their
 neutron-capture elements nature}. The first sub-class is the CEMP-s, where stars show high enhancements
 in carbon, together with enhancements in elements formed mainly by the slow (s-) neutron capture process
 (\citealt{2013ApJ...770..104P,  2015A&A...581A..22A}).
  The second sub-class is CEMP-r/s, where stars show enhancements in 
   both slow (s-) and rapid (r-) process material. 
 The studies of \citet{2005ApJ...625..825L}, \cite{2014MNRAS.441.1217S} and \cite{2016A&A...588A...3H} 
 confirmed the binarity of CEMP-s stars, so their peculiar chemical pattern can be explained by pollution from 
 an asymptotic giant branch (AGB) star companion, which has since become a white dwarf. The wide variety 
 enhancement associated with CEMP-r/s can not be explained by the binarity of involving AGB stars 
 \citep{2005ARA&A..43..531B, 2007ApJ...655..492A}.
 
  Recently, the Laser Interferometer Gravitational Wave Observatory (LIGO) and Virgo Collaboration (LVC) 
  were able to detect the loudest gravitational wave (GW170817), generated by
   binary neutron star merger \citep{2017PhRvL.119p1101A}. \citet{2017Sci...358.1574S} studied the optical 
   counterpart of GW170817 namely Swope Supernova Survey 2017a (SSS17a), where they showed 
   (in Figure 4A) that SSS17a is consistent with the expectations for r-process heating, which supports the idea
    that the main site of the r-process production is the neutron star mergers  \citep{2007PhR...442..237Q, 
    2007PhR...450...97A, 2017Natur.551...71T}. A few CEMP stars {exhibit overabundances} of the 
 elements formed mainly by the r-process \citep{1957RvMP...29..547B}. 
 In all probability, {their r-process enhancement took place in their birth cloud}, when it is enriched by r-process material. 
 These stars are known as CEMP-r. The r-process-rich ejecta mix with the interstellar medium (ISM), which will become star-forming regions 
  and form new generations of r-rich stars \citep{2016Natur.531..610J}. 
  
  {Recently, it has been recognized that the majority of CEMP stars at [Fe/H] $\leq -4$ show no 
  enhancement in neutron capture elements} (e.g., \citealt{2002Natur.419..904C, 2005Natur.434..871F, 
2011Natur.477...67C, 2014Natur.506..463K}) . These stars are classified as CEMP-no. While CEMP-s
  were have been established as member of binary systems, CEMP-no were found to be inconsistent with the binary
  properties of the CEMP-s class \citep{2014MNRAS.441.1217S,2016A&A...588A..37H}, thereby strongly indicating
  a different physical origin of their carbon-enhancement. These stars most likely appear to exhibit abundance patterns
  of its natal cloud, that still makes their unusual chemical composition a puzzle \citep{2013ApJ...773...33I, 
  2014ApJ...797...21P, 2015ApJ...812..109P, 2016ApJ...829L..24P, 2016ApJ...821...37R, 2016ApJ...833...20Y}.

{In addition to carbon enhancement, rare CEMP stars exhibit nitrogen overabundances, typically 
with [N/Fe] $>$ 1.0 and [N/C] $>$ 0.5, although this phenomenon depends on metallicity, it appears to be more 
frequent at [Fe/H] $<2.8$ ($\sim$ factor of 10) \citep{1982ApJ...263L..29B, 2007ApJ...658.1203J, 
2009PASA...26..327P, 2012A&A...547A..76P, 2014ApJ...784..158R}}.
The existence of a population of nitrogen-enhanced metal-poor (NEMP) stars was predicted from the same evolution scenario,
since AGB stars efficiently cycle carbon into nitrogen in their envelopes.
Detailed AGB nucleosynthesis models of low initial mass ($<$2.5 M$\odot$) produce carbon, 
but do not produce nitrogen because it is burned during helium shell flashes. On the other hand, AGB models of higher 
mass convert the dredged-up carbon into nitrogen by CN cycling at the bottom of the convective envelope (hot bottom burning, HBB)
\citep{2005A&A...437..429C, 2007A&A...461..571H, 2008A&A...489..685E, 2009A&A...508.1359I, 2010ApJ...709...11J}.

Metal poor stars reside primarily in the halo system of the Milky Way, a complex old component of the Galaxy 
that comprises at least two diffuse stellar populations, the inner- and the outer-halo, with different metallicities,
kinematics and spatial density profiles \citep{2007Natur.450.1020C, 2010ApJ...712..692C, 2012AAS...21922206B}, 
several streams and overdensities \citep{2009ApJ...693.1118G}, and a recently discovered large structure in the 
inner region, product of a past merger event \citep{2018Natur.563...85H}. It has been recognized that 15\% $-$ 20\% 
of stars with [Fe/H] $< -$2.0 in the halo system are CEMP and the fraction increases with declining metallicity becoming 
$\sim$ 75\% at [Fe/H] $< -$4.0  \citep[see][and reference therein]{2014ApJ...788..180C}. There is also evidence for a 
significant contrast in the frequency of CEMP stars that are kinematically assigned to the inner- and outer-halo 
components (the inner halo is on-average non rotating, while the outer halo exhibits a significant retrograde signature). 
The outer halo exhibits a fraction of CEMP stars twice the inner halo in the metallicity interval $-$2.5 $<$[Fe/H]
 $-$2.0 \citep{2012ApJ...744..195C}. Such increase in frequency of CEMP stars can be explained as a population driven effect, 
 due to the fact that the outer halo is the dominant component at large distance from the galactic plane and at metallicities, 
 [Fe/H] $< -$2.0. The chemical differences between the two stellar halo populations were established also in terms of CEMP 
 sub-classes by \citet{2014ApJ...788..180C}. It was shown that the relative numbers of CEMP-no stars compared to CEMP-s 
 varies between the inner- and outer-halo and the frequency of the CEMP-no stars is higher in the outer halo, while the 
 frequency of the CEMP-s stars is higher in the inner halo. The analyses of kinematics and dynamics of our sample stars will 
 establish their inner/outer halo membership and on their origin.

The chemical abundance analysis of metal-poor stars and searching of CEMP and NEMP stars provides important constrains
on the chemistry evolution of the Galaxy, initial mass function (IMF; \citealt{2014ApJ...781...60H}) 
and the models of mass-transfer and evolution of components in binary systems. In this paper, 
we report on the discovery of {five CEMP stars}, which add impact constraints on different stellar and 
Galactic chemical-evolution scenarios, as well as the nature of their progenitors.

This paper is outlined as follows: target selection and observations are discussed in Section \ref{sect:obs}.  
Section \ref{sec:three} discusses the determination of stellar parameters. Our abundance analysis 
is discussed in Section \ref{sect:Method}. We discuss our results, chemical peculiarities, and kinematics
in Section \ref{sec:discussions}, and our conclusions are given in Section \ref{sect:conclusion}.

\section{Target Selection, Observations and Data Reduction}   
\label{sect:obs}
\subsection{LAMOST database}
Our sample was selected from the Large Sky Area Multi-Object Fiber Spectroscopic Telescope survey
\citep{2006ChJAA...6..265Z, 2012RAA....12..723Z, 2012RAA....12.1197C}, where we ran two different methods independently in order to 
estimate the total metallicity [Fe/H] of an object. These two methods can be highlighted by the following points:

\begin{itemize}
\item Matching synthesized spectra {with} the observed data, and calculate the corresponding 
Lick indices to find the best parameters \footnote{See http://astro.wsu.edu/worthey/html/index.table.html}.
\item Matching observed normalized spectra with synthesized spectra using a minimization technique to carry out the best-fit for metallicity.
\end{itemize}

After running these two methods, {we identified a star} as very metal-poor (hereafter VMP) candidate,
if these two methods yield [Fe/H] $\leq -2.7$, assuming a typical uncertainty of deriving metallicities from low-resolution spectra ($0.1$-$0.3$ dex).

\subsection{High resolution spectroscopy}
We {obtain high-resolution} spectra (R = 110,000 and a slit width of 0.5 mm), covering the 
wavelength range of ($3730$-$9989$ {\AA}\,), {for a sample of 12 stars}, using the 2.4m Lick Automated Planet Finder (thereafter Lick/APF).
In addition, we observed the well studied VMP star HD2796 as a standard.
For more information about this telescope and its 
instruments, we refer the reader to \citet{2014SPIE.9145E..2BR}. 
After removing the dispersion caused by atmospheric refraction 
\footnote{This step was achieved using a trombone-style atmospheric dispersion corrector (ADC)}, we carried out a standard echelle data reduction 
(bias subtraction, flat- fielding, background subtraction, extraction, wavelength calibration and continuum-normalization) to obtain 1D normalized spectra. 
During our analysis we found that {five} stars showed high carbon abundances. In this paper, we present these {five} stars and in a future paper we will present the remaining sample.
Table \ref{tab:obs} shows the observational details of our sample. The seventh column lists signal-to-noise ratios (S/N), measured using IRAF splot task at 
$\sim$ 4500{\AA}\,. {For the radial velocity (RV) measurements, we built a routine to estimate the heliocentric corrections of our observations and 
cross-correlate our spectra against synthesized templates, with the same spectral type of each star, using strong features (e.g., Mg I  triplet).
In addition, we checked the validity of our method using some metal-poor radial velocity standards. These values are also listed in the last column of Table \ref{tab:obs}}.

\subsection{Equivalent Widths}\label{sec:EWs_uncer}

We adopted an atomic line-list from \citet{2013AJ....145...13A} and
\citet{2013ApJ...769...57F} (also see Section \ref{sub:carb-nitro}), we used Gaussian profiles to measure 
the equivalent widths (thereafter EWs) of those isolated atomic lines \footnote{Gaussian profiles can not fit strong lines very well, but still lines with 
EWs $\sim$ 100m{\AA}\, are still gaussian-like shapes.}. Section \ref{sect:Method} describes
in more details our abundance measurement method. The measured EWs are listed in Table \ref{tab:EWs}.

\begin{figure}
\epsscale{1.2}
\plotone{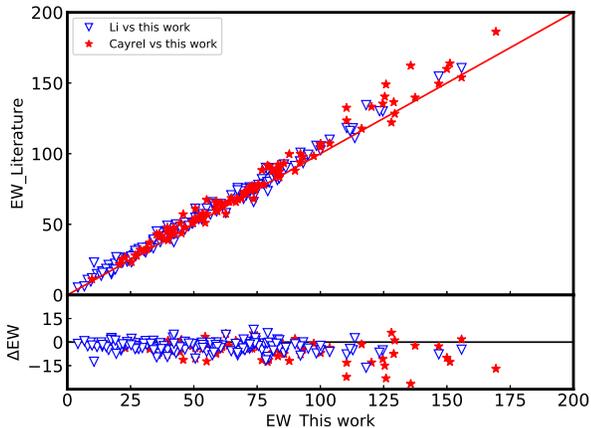}
\caption{Comparison of the equivalent widths of HD2796, the blue circles represent 
the comparison of this work and \citet{2004A&A...416.1117C} ,while the black triangles represent the comparison of this this work and  \citet{2015ApJ...798..110L}. {Note that the biggest discrepancies between our EWs and the literature data are for the strongest lines. Thus, the spectrum synthesis approach have been used for these lines (e.g., Mg I  tripletsg)}.
\label{fig:EWsquality}}
\end{figure}

In order to assess the quality of our EWs measurements 
we compared our EWs of HD2796 with \citet{2004A&A...416.1117C}
and \citet{2015ApJ...798..110L}, shown in Figure \ref{fig:EWsquality}. Our EWs have good agreement with Cayrel's results 
($R^{2}=0.9806$, zero-point shift $-0.107$ and $\sigma$ of $3.1$ m{\AA}\,), and Li's results ($R^{2}=0.9879$, 
zero-point shift +1.0234 and $\sigma$ = 2.1 m{\AA}\,). The upper panel of Figure \ref{fig:EWsquality} shows a direct comparison
of our EWs and their measured values for the common lines, and one-to-one line is used as a reference. The lower panel shows
very small residuals of the EWs (this work - literature)  for those $100$ m{\AA}\,$>$EW$> 10$ m{\AA}\,.

Neglecting the uncertainties of our continuum definition and based on \citet{1988IAUS..132..345C} formula (see Eq. \ref{eq:1}), 
we expected our uncertainties to be around 2.5 m{\AA}\,. 
However, one should note that those values are for the weakest detected lines in these spectra. 
Where, $FWHM$ is the full width at half maximum,  $S/N$ is the signal-to-noise ratio and $\delta x$ is the pixel size.

\begin{equation}\label{eq:1}
\sigma_{EW} = \frac{1.5}{S/N}~\sqrt{FWHM*\delta x}
\end{equation}

\section{Stellar parameters}\label{sec:three}
Despite the fact that iron abundances derived from both excitation levels (\ion{Fe}{1} 
and \ion{Fe}{2}) are affected by uncertainties in the model atmospheres temperature and
NLTE effects \citep[see][and reference therein]{2017A&A...604A.129M}, many researches still rely on this method (the so-called traditional 
spectroscopic method) to derive their stellar atmospheric parameters. 
Our adopted stellar parameters have been determined through this standard spectroscopic method. 
Additionally, we were able to determine effective temperatures from photometry and surface gravities from parallax/distances.

\subsection{Effective temperature}
The effective temperatures (\Teff) were estimated by 
minimizing the trend between \ion{Fe}{1} lines abundances and excitation potential ($\chi$). 

To estimate \Teff~from the available colors (B,V,J,H and K), we cross match our sample with two catalogs 
from the Virtual Observatory:  The fourth US Naval 
Observatory CCD Astrograph Catalog (UCAC4, \citealt{2013AJ....145...44Z} )
and Two Micron All Sky Survey (2MASS, \citealt{2006AJ....131.1163S}) and then
employ \citet{2005ApJ...626..465R} temperature calibration.
{These \Teff~(determined from photometric and spectroscopic methods) together with \Teff~
taken from Gaia DR2 \citep{2018A&A...616A...1G} are in good agreement with each other ($\pm 150$ K)}.

\subsection{Surface gravity and microturbulence}
We determined the surface gravity \ensuremath{\log\,g}) by forcing \ion{Fe}{1} abundances to 
agree with \ion{Fe}{2} abundances. In addition, we crossed match our sample with Gaia DR2 distances catalogue \citep{2018AJ....156...58B}, to estimate 
the distance modulus from the available parallax of \citet{2018A&A...616A...1G} (see Eq. \ref{eq:2} and Eq. \ref{eq:3}).

\begin{equation}\label{eq:2}
\log \frac{g}{g_{\odot}} = \log \frac{M}{M_{\odot}} + 4 \log 
\frac{T_{\rm eff}}{T_{\rm eff}} + 0.4(M_{bol} - M_{bol_{\odot}})
\end{equation}

\begin{equation}\label{eq:3}
M_{bol} = V+BC+5 \log \varpi+5
\end{equation}

\begin{figure}[t!]
\epsscale{1.2}
\plotone{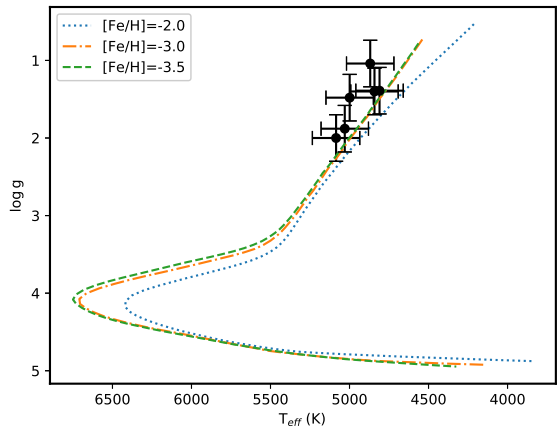}
   \caption{Our sample shown in an H-R-Diagram, based on the \Teff ~and \ensuremath{\log\,g}  determined from the LIKC/APF spectra (see Table \ref{tab:stellar-param}).
   Yale-Yonsei 12 \Gyr~ isochrones with [$\alpha$/Fe] = +0.4 and [Fe/H] = $-$2.5, $-$3.0, and $-$3.5 from \citet{2004ApJS..155..667D}  overplotted as reference.}
\label{fig:isoch}
\end{figure}

where, $M$ is the stellar mass, $M_{bol}$ is the absolute bolometric magnitude, 
$V$ is the visual magnitude, $BC$ is the bolometric correction \citep[see][Eq. 18]{1999A&AS..140..261A}, and $\pi$ is the parallax.

The microturbulence velocities ($\xi$) were determined by removing any trend between \ion{Fe}{1} lines 
abundances with the EWs of those lines.

{The \Teff\,determined from the photometric and spectroscopic methods show 
systematically different results. \citet{2013ApJ...769...57F} have presented an explicit method to adjust the spectroscopic \Teff. 
This scheme increases the \Teff\,determined for cool red-giants up-to several hundred degrees, 
on the other hand the \Teff\,determined for main-sequence stars are mostly unaffected.
Motivated by \citet{2013ApJ...769...57F} results, the atmospheric stellar parameters 
(\Teff, \ensuremath{\log\,g} and $\xi$) determined for our sample stars from the spectroscopic method were considered 
as initial parameters, and then corrected following the same scheme presented in \citet{2013ApJ...769...57F}.
It's worthy to note that our parameters were not determined independently. Thus, this procedure was iterated to consistency.
The derived stellar atmospheric parameters are listed in Table \ref{tab:stellar-param}.}

The corrected spectroscopic surface gravities (in cgs units) 
versus the corrected spectroscopic \Teff~of our programme stars, with 12 \Gyr~Yale-Yonsei isochrones as a reference \citep{2004ApJS..155..667D} 
are shown in Figure \ref{fig:isoch}. {The error bars shown in Figure \ref{fig:isoch} represent \Teff~ and \ensuremath{\log\,g} one-sigma errors ($\pm 150$ K and $\pm 0.3$ cgs, respectively).}

\section{Abundance Analysis}
\label{sect:Method}

We used only non-blended lines with reliable continuum normalization to measure the chemical abundances using EWs analysis.
However, for the molecular bands and blended lines, spectral synthesis was used. 
In other words, our chemical abundances were done by a mixture of spectrum synthesis 
and equivalent width analysis. The LTE abundances for all elements are listed in Table \ref{tab:res}.
Moreover,  we considered the deviations from LTE for \ion{Li}{1} , \ion{Na}{1} and \ion{Mg}{1} lines. 
The LTE and NLTE abundances are presented in Table \ref{tab:nlte}. 
We adopted the solar log$\epsilon_\odot$(X) from \citet{2009ARAA..47..481A} to obtain our final chemical abundances and [X/Fe] ratios.

\begin{figure}[t!]
{\includegraphics[width = 1.5in]{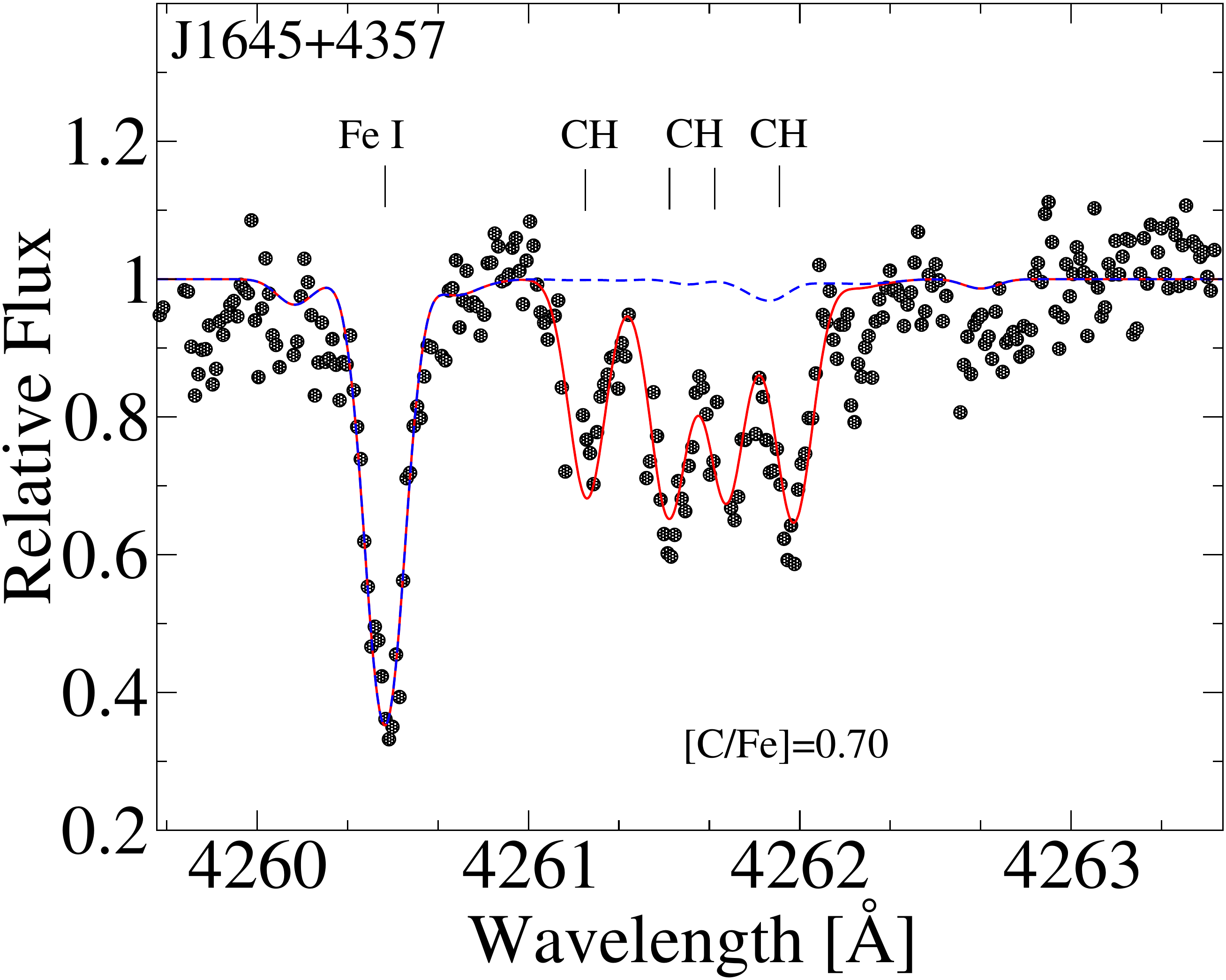}} 
{\includegraphics[width = 1.5in]{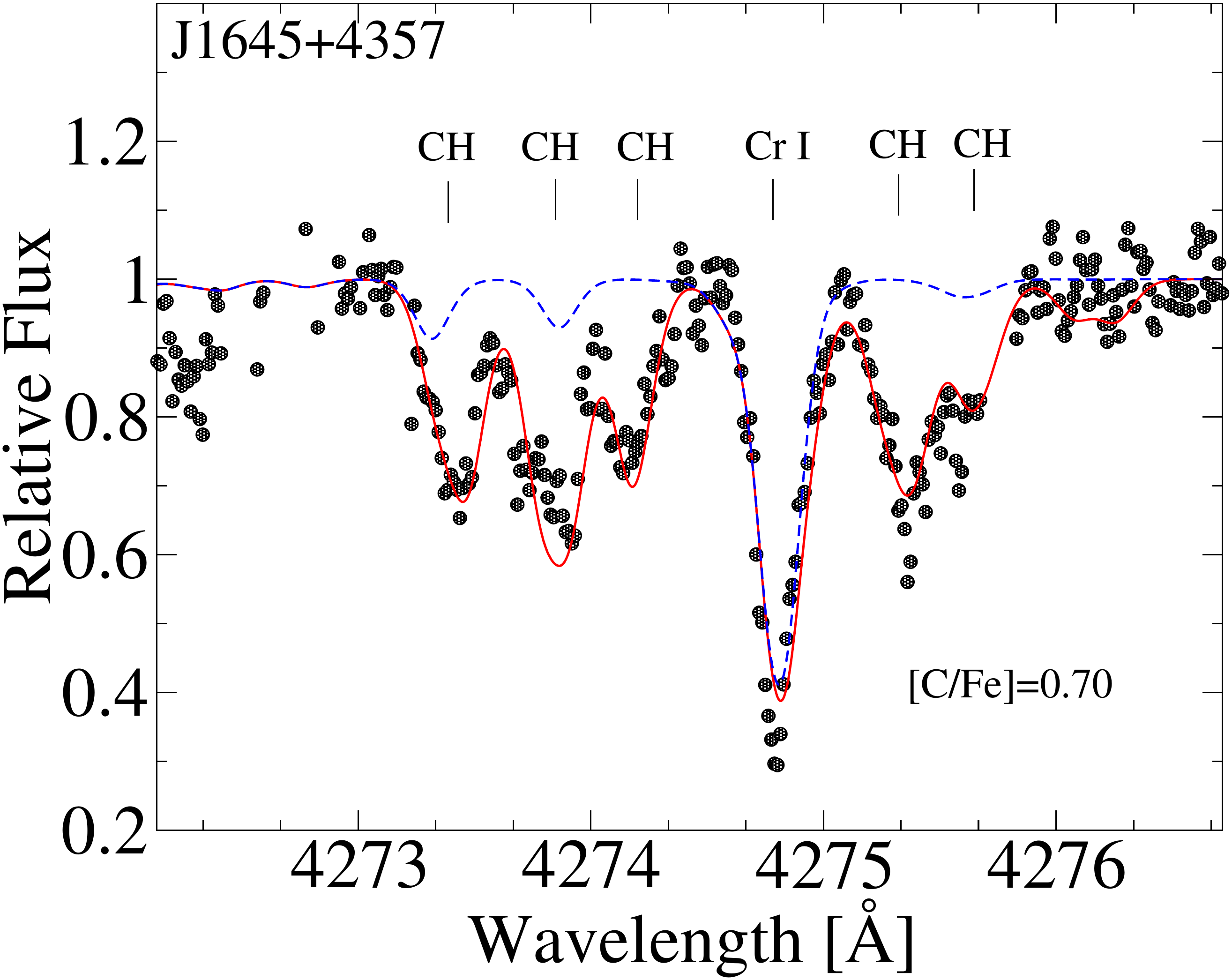}}\\
{\includegraphics[width = 1.5in]{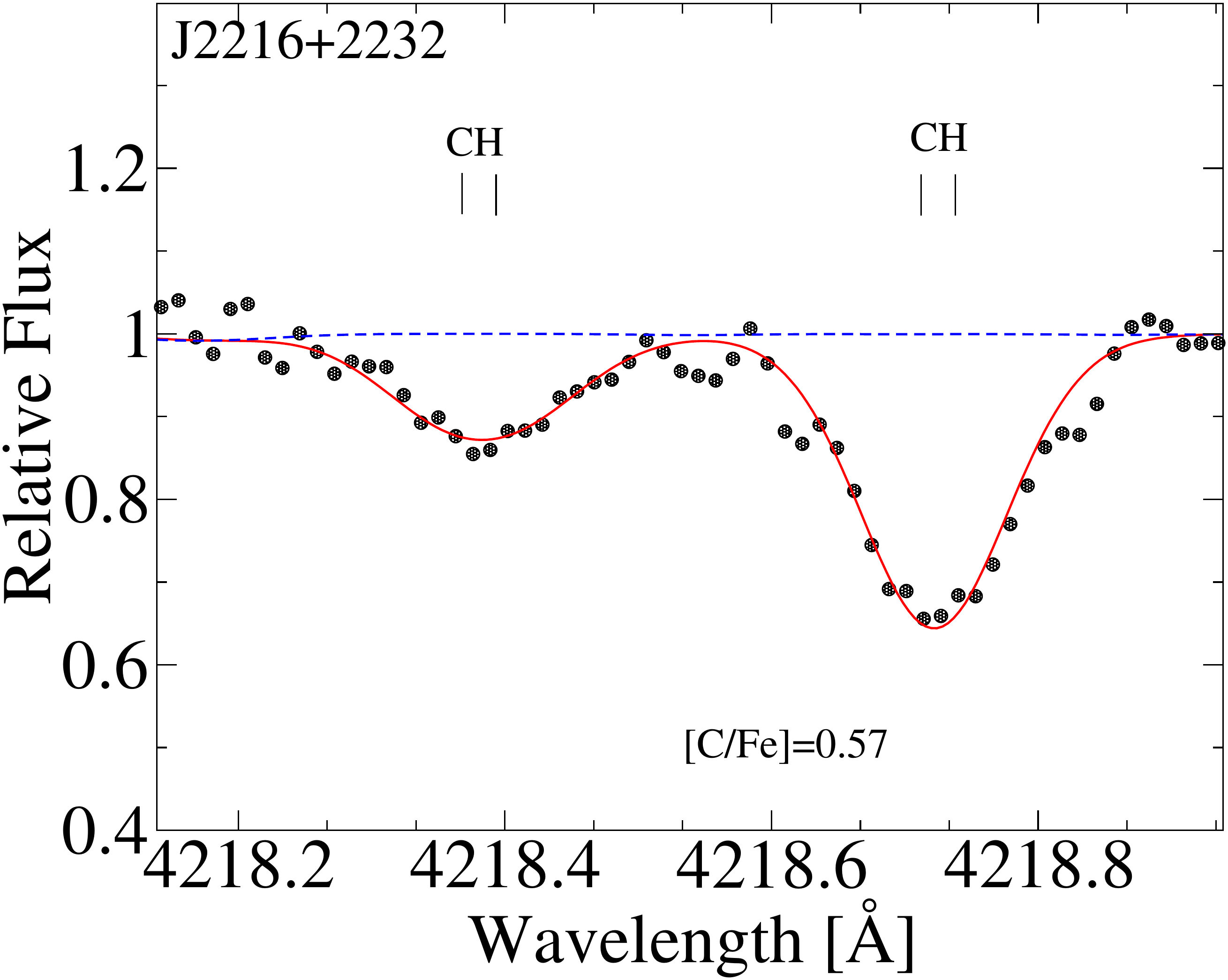}}
{\includegraphics[width = 1.5in]{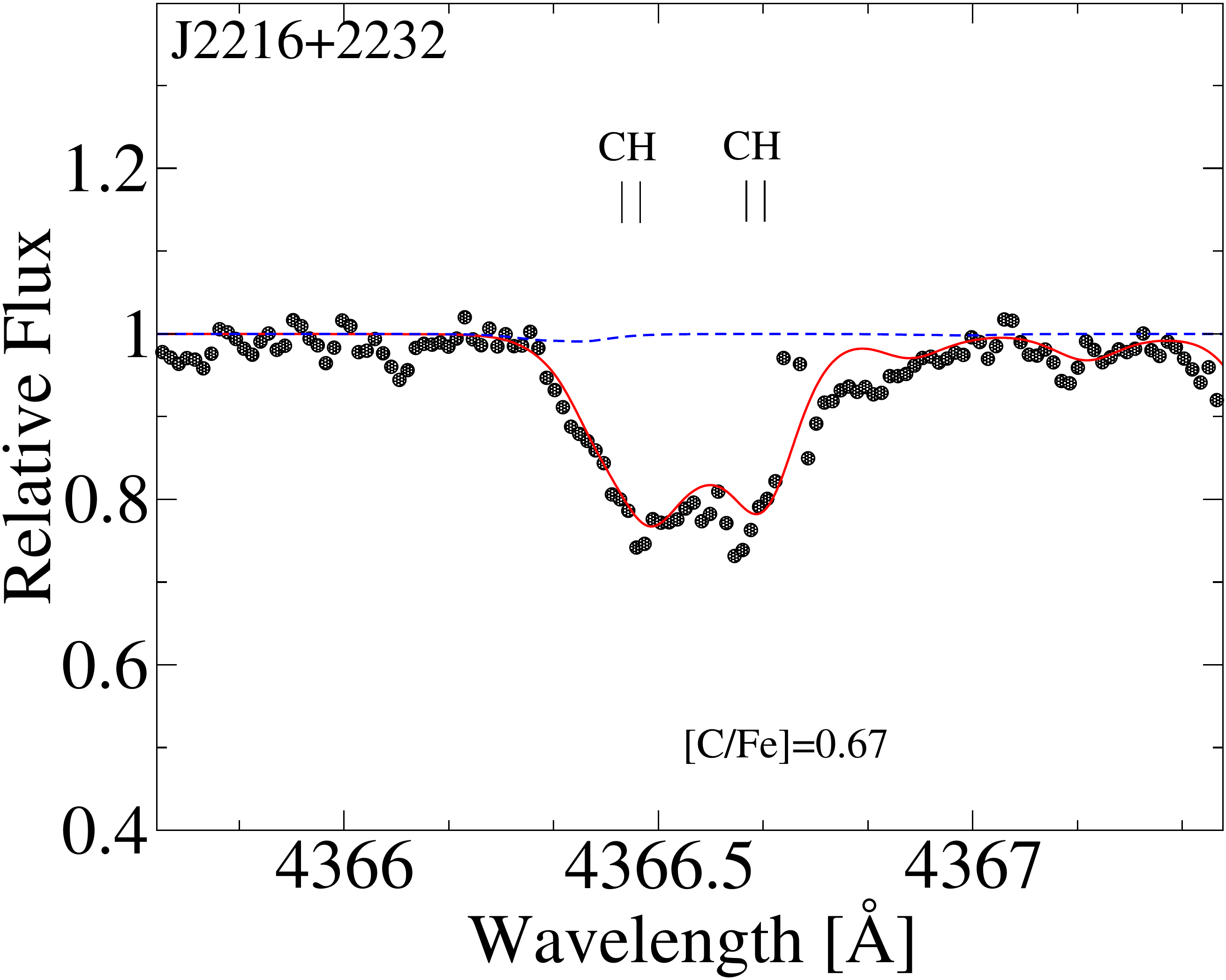}} \\
{\includegraphics[width = 1.5in]{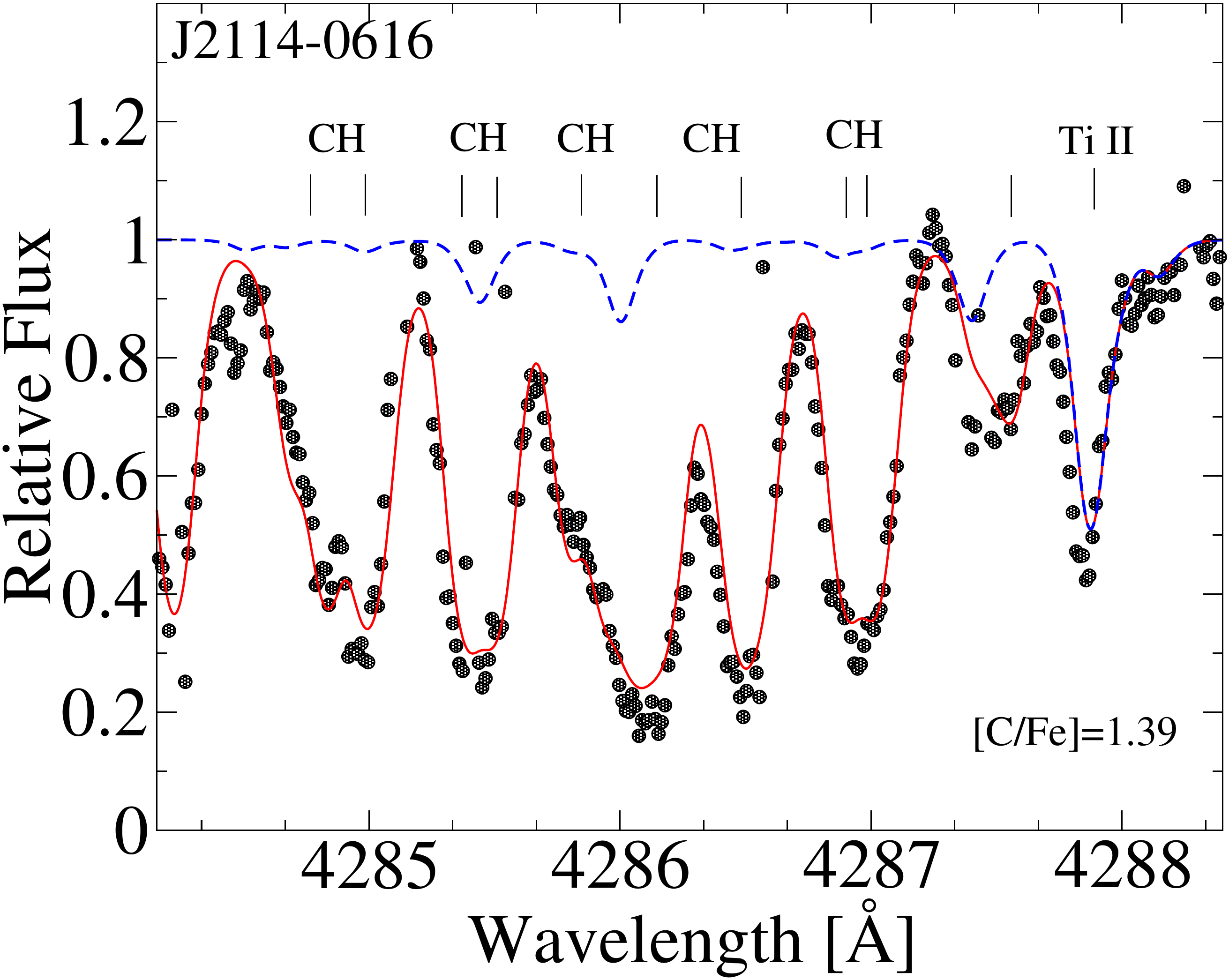}}
{\includegraphics[width = 1.5in]{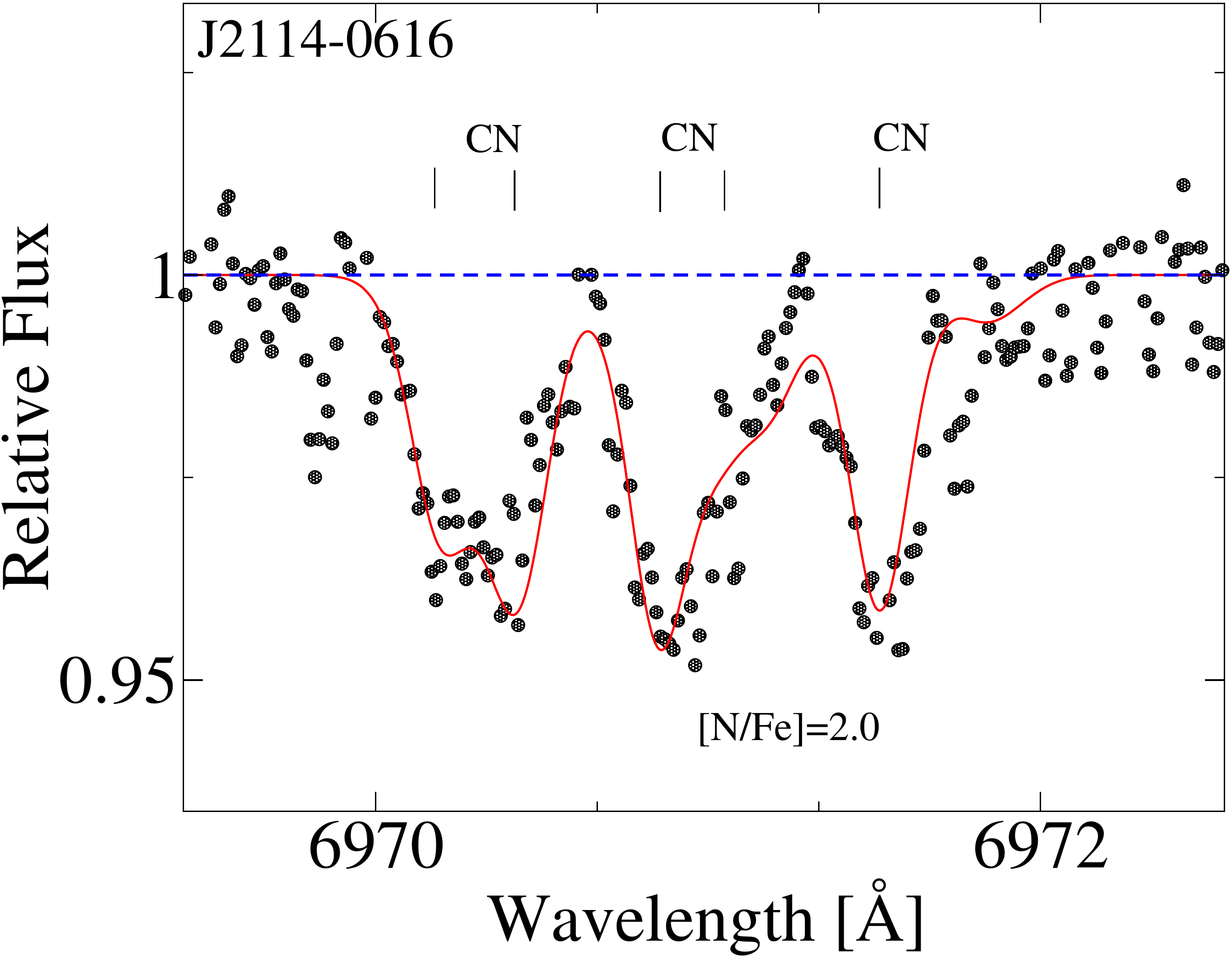}}
  \caption{Portions of the LICK/APF spectra near the CH and CN bands (are represented as black filled circles). Best-fit of the selected CH and CN are also shown (red solid line).
  The dashed lines show the synthetic spectrum with no carbon in the atmosphere.}
 \label{fig:best-fit-lines}
  \end{figure}  

\subsection{LTE and NLTE calculations}

We used stellar atmosphere models from 1D ATLAS NEWODF grid of \citet{2003IAUS..210P.A20C}.
Our LTE abundances were performed using an updated version of the stellar code MOOG \citep{1973PhDT.......180S}.
In this update, a continuous scattering will be treated as a source function, in other-words the absorption and scattering
will be summed rather than treated as true absorption \citep{2011AJ....141..175S}.

The departures from LTE in the stellar atmospheres were considered for three chemical elements, Li, Na, and Mg. 
The adopted Na I and Mg I-II model atoms are described in \citet{2014AstL...40..406A} and \citet{2018arXiv180906969A}, respectively.
To solve the radiative transfer and statistical equilibrium equations, we used the code \textsc{detail} \citep{detail} based on the accelerated
 $\Lambda$-iteration method \citep{rh91}. The obtained departure coefficients, $b_{\rm{i}}$ = $n_{\rm{NLTE}}$ / $n_{\rm{LTE}}$, were then 
 used by the codes \textsc{binmag3} \citep{binmag3} and \textsc{synthV-NLTE} \citep{2016MNRAS.456.1221R} to calculate the synthetic NLTE line profiles. 
Here, $n_{\rm{NLTE}}$ and $n_{\rm{LTE}}$ are the statistical equilibrium and thermal (Saha-Boltzmann) number densities, respectively. 

We constructed the \ion{Li} {1} model atom in the same manner as it was described in \citet{2009A&A...503..541L}. 
The main difference between our model atom and the model atom of \citet{2009A&A...503..541L} is the collision excitation recipe. 
We adopted the electron collision data from \citet{2011A&A...529A..31O}, while 
\citet{2009A&A...503..541L} used cross-sections for collisional excitation by electrons from \citet{1971JQSRT..11....7P}. 
We tested our model with Li-enhanced stars and have found a good agreement with \citet{2009A&A...503..541L}.

\subsection{Lithium}\label{sec:lith}
Our LICK/APF spectra showed no obvious features for the lithium abundance determinations, except in J2216$+$2232 spectrum.
{Therefore, upper-limits were determined for the rest of the sample stars.}
The lithium abundance was derived from the \ion{Li}{1} 6707.7 {\AA}\, resonance line. The line-list was taken from VALD database, 
hyperfine structure and isotope structure taken into account with the data from \citet{1995PhRvA..52.2682S}.

\subsection{Carbon and Nitrogen}\label{sub:carb-nitro}

\textbf{Our carbon abundances were derived from the molecular CH AX band around 4300 {\AA}\,}
The molecular line data for the spectrum synthesis was taken from VALD \citep{vald} database. 
{We determine carbon abundances using the method described in \citet{2015MNRAS.453.1619A}}, 
our synthetic flux profiles were convolved with a profile that combines a rotational broadening and broadening by
macroturbulence with a radial-tangential profile. 
The most probable macroturbulence velocity V$_{mac}$ was varied between 4 and 9 \kms for different CH lines.

The nitrogen abundances were estimated from the CN 4215 {\AA}\, and 6971{\AA}\, bands, using a spectrum synthesis approach. 
The dissociation energy (D$_0$) of CN was adopted to be 7.65 eV \citep{1988ApJ...332..531B}.
Only in J2114$-$0616 spectrum, we have found visible CN 4215 {\AA}\, and 6971{\AA}\, bands, which can be measured quite reliably.
The best fits of some molecular lines in our sample are shown in Figure. \ref{fig:best-fit-lines}.

\subsection{Light Elements: from Na to Zn}

\begin{figure}[t!]
{\includegraphics[width = 1.5in]{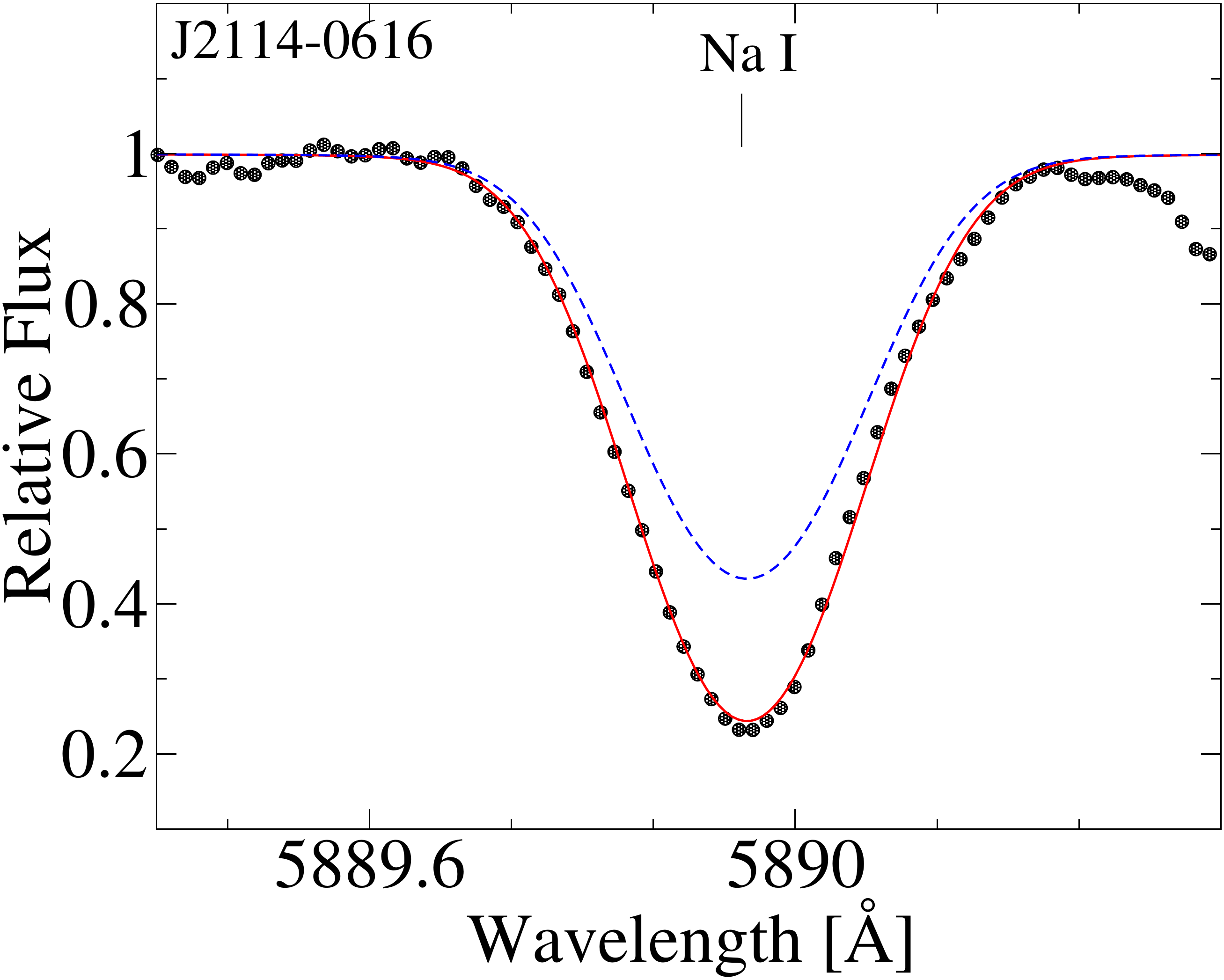}} 
{\includegraphics[width = 1.5in]{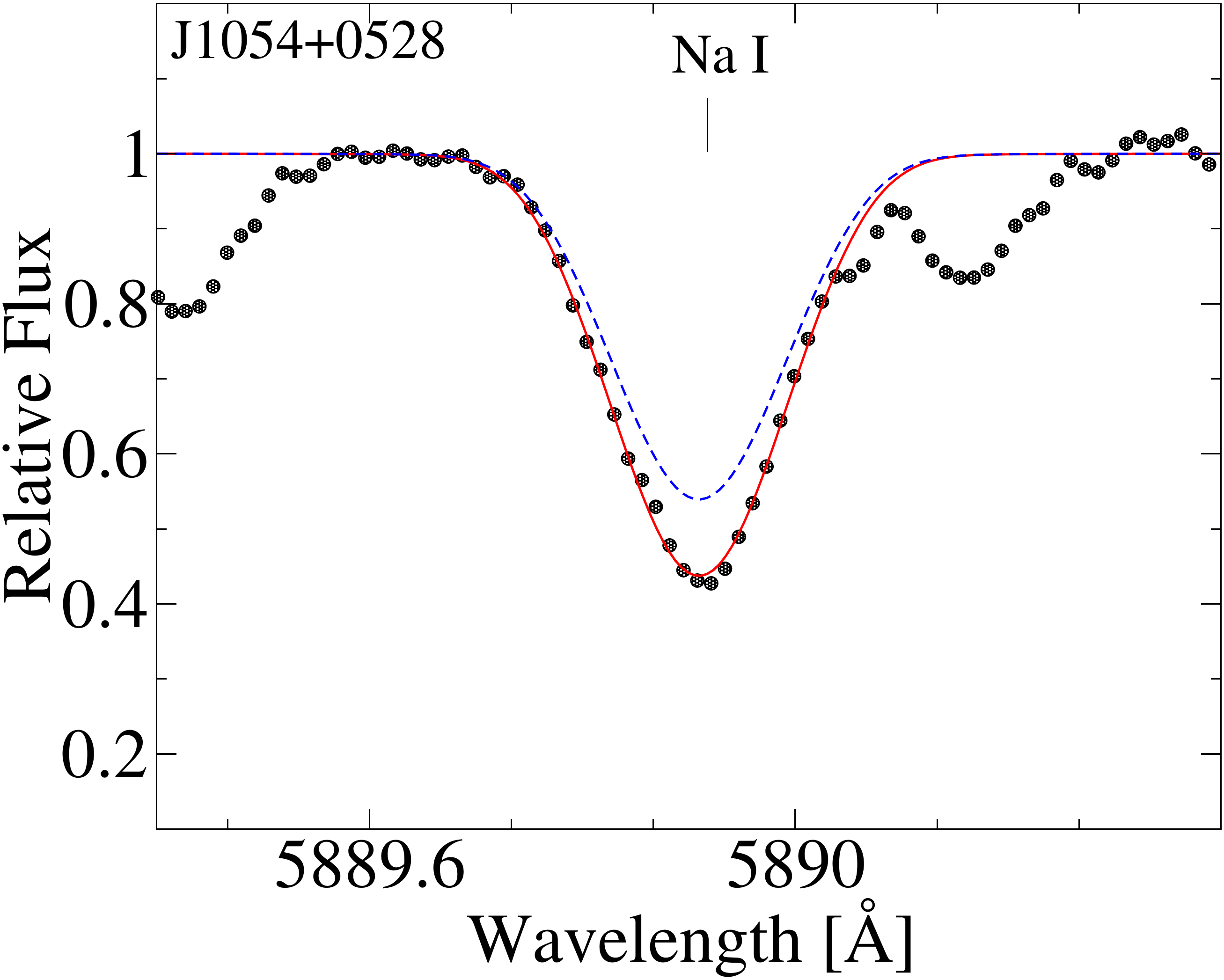}}\\
{\includegraphics[width = 1.5in]{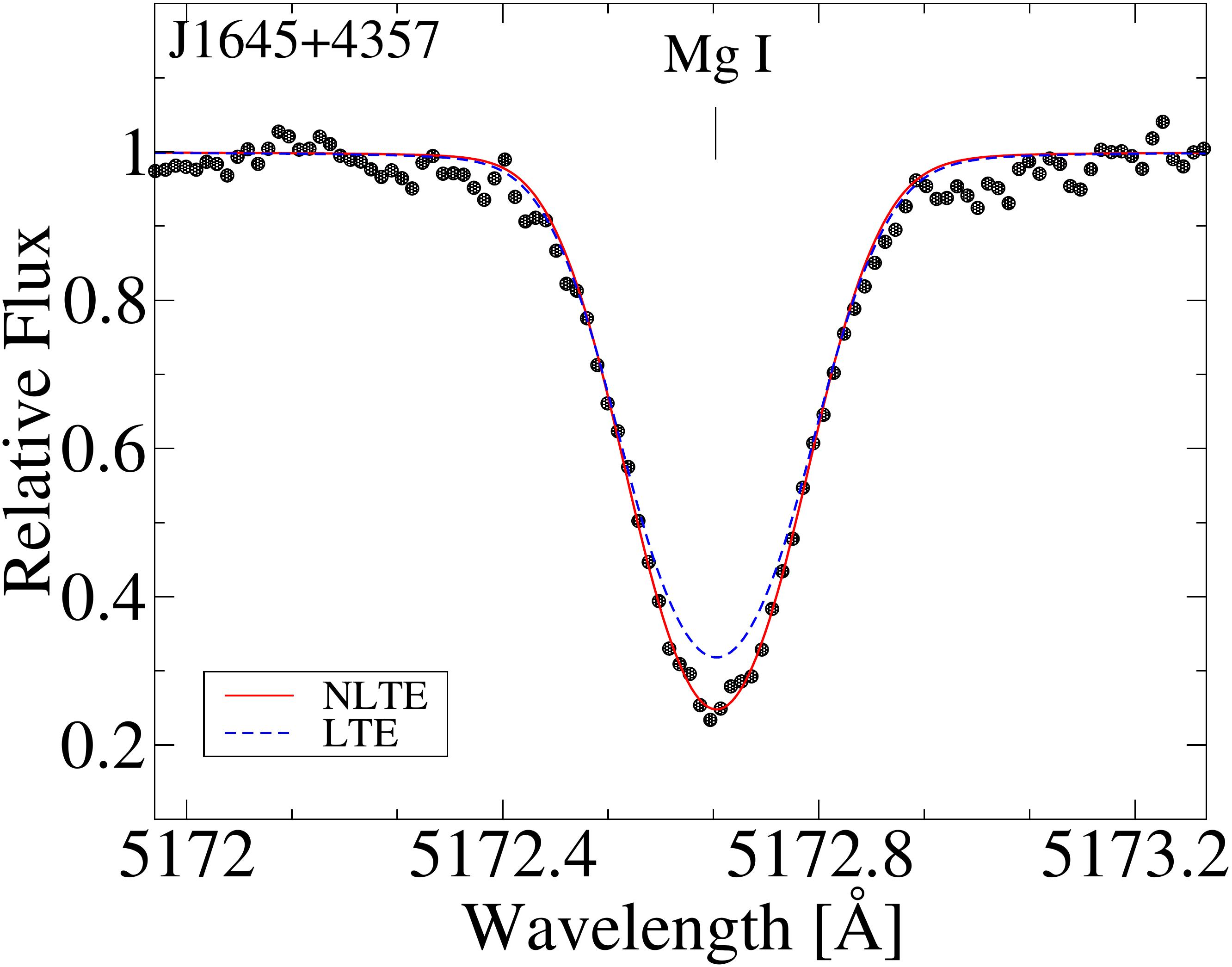}} 
{\includegraphics[width = 1.5in]{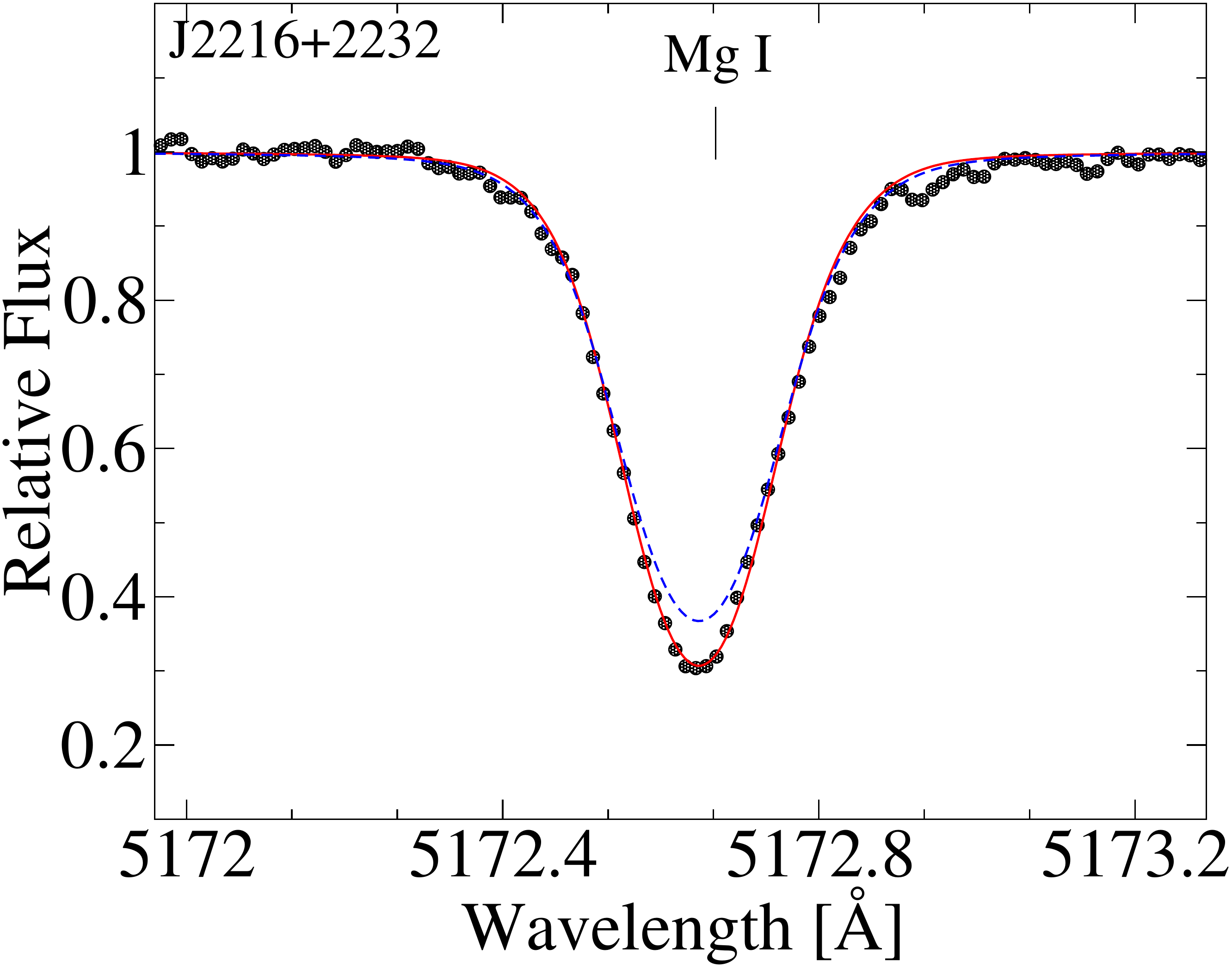}}\\
\caption{An example of the NLTE fittings of Na and Mg in LICK/APF spectra . The observed spectra are shown by filled circles, for comparison, the LTE profiles computed with the corresponding NLTE abundances are shown by the blue dashed curves. }
\label{fig:Na-best}
\end{figure}

We were able to measure abundances for Na, Mg, Ca, Ti, Sc, V, Cr, Mn, Co, Ni, and Zn in the LICK/APF spectra.
Our sample stars are metal-poor and only resonance Na I lines at 5889, 5895 {\AA}\, are available for measurements. 
The van der Waals damping constant, C$_6$ = $-$31.6, for these lines was adopted from solar line-profile fitting \citep{2016ApJ...833..225Z}. 
The magnesium abundances are derived from Mg I lines at 4703, 5172, 5183, 5528, 5711 {\AA}\, (see Figure \ref{fig:Na-best}).

\begin{figure}
\epsscale{1.2}
\plotone{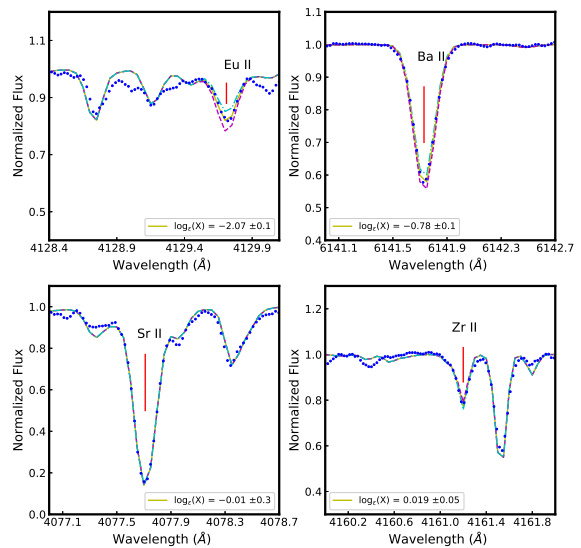}
   \caption{Portions of the LICK/APF spectrum of HD2796 (shown as blue  filled circles) near the lines of \ion{Eu}{2} at 4129  {\AA}\, (top left),  \ion{Ba}{2} at 6141  {\AA}\, (top right), \ion{Sr}{2} at 4077 {\AA}\,(bottom left), and \ion{Zr}{2} at 4161 {\AA}\,(bottom right). Best-fit synthetic spectra are also shown (yellow solid line), together with abundance variations (magenta dashed line and cyan dash-dotted line) of $\pm$0.1 dex (for Eu), $\pm$0.1 dex (for Ba), $\pm$0.3 dex (for Sr), and $\pm$0.5 dex (for Zr).}
\label{fig:ncapture}
\end{figure}

For the most of our sample stars, Ca abundances were derived from 18 well-defined \ion{Ca}{1} lines, Ti 
abundances were obtained from 26  \ion{Ti}{1} and \ion{Ti}{2} lines in total, 
Scandium appeared in more than six \ion{Sc}{2} lines, while V has only one reliable line at \ion{V}{1} 4379.23 {\AA}\,.

\ion{Cr}{1} has at least 5 lines, while \ion{Mn}{1} and \ion{Sc}{2} lines appeared in all sample spectra. 
In our standard star HD2796 spectrum we were unable to find any detectable feature 
for \ion{Co}{1} and \ion{Ni}{1}, while they appeared in the remaining stars of our sample. Moreover, J1645$+$4357 spectrum 
shows no detectable line for the heaviest element of the iron peak Zn, see Table \ref{tab:res}.

\subsection{Neutron-capture Elements}

The strontium (Sr) and barium (Ba) abundances for metal-poor stars are very important, since they are the most commonly detected neutron-capture elements
whose abundances are measured in the vast majority of the metal-poor stars, thus these two elements are the key of understanding the nature of the neutron-capture processes in our Galaxy.

One of our main aims of the LICK/APF observations was to increase the high-resolution chemical inventory of metal-poor stars.
In addition, to these elements mentioned previously, we also investigated yttrium, zirconium, lanthanum, cerium, 
praseodymium, neodymium, samarium, and europium, see Table \ref{tab:res}. 

Figure \ref{fig:ncapture} shows an illustrative spectrum-synthesis example of the neutron-capture elements in our standard star HD2796.
These elements show very weak features compared to those elements with Z$\leq30$,
with abundance uncertainties varying between 0.1 and 0.3 dex, depending on the quality of our spectrum-synthesis abundance measurements.

 \subsection{Uncertainties of stellar parameters and abundances}\label{subsec:abuns_uncer}
In Section \ref{sec:EWs_uncer} we discussed the uncertainties arising from the EWs. The stellar parameters are subject of another 
uncertainties arising from our ionization equilibrium process (the traditional spectroscopic method), where we expect that this process will provide 
typical internal accuracy in our estimated surface gravity and microturbulence of 0.1 dex and 0.2 km s$^{-1}$  respectively. In addition, these 
uncertainties will affect our abundances determination. The major uncertainties in our abundance determinations are related to the 
estimated effective temperatures (usually in the order of 100 K).  

\begin{figure}
\epsscale{1.0}
\plotone{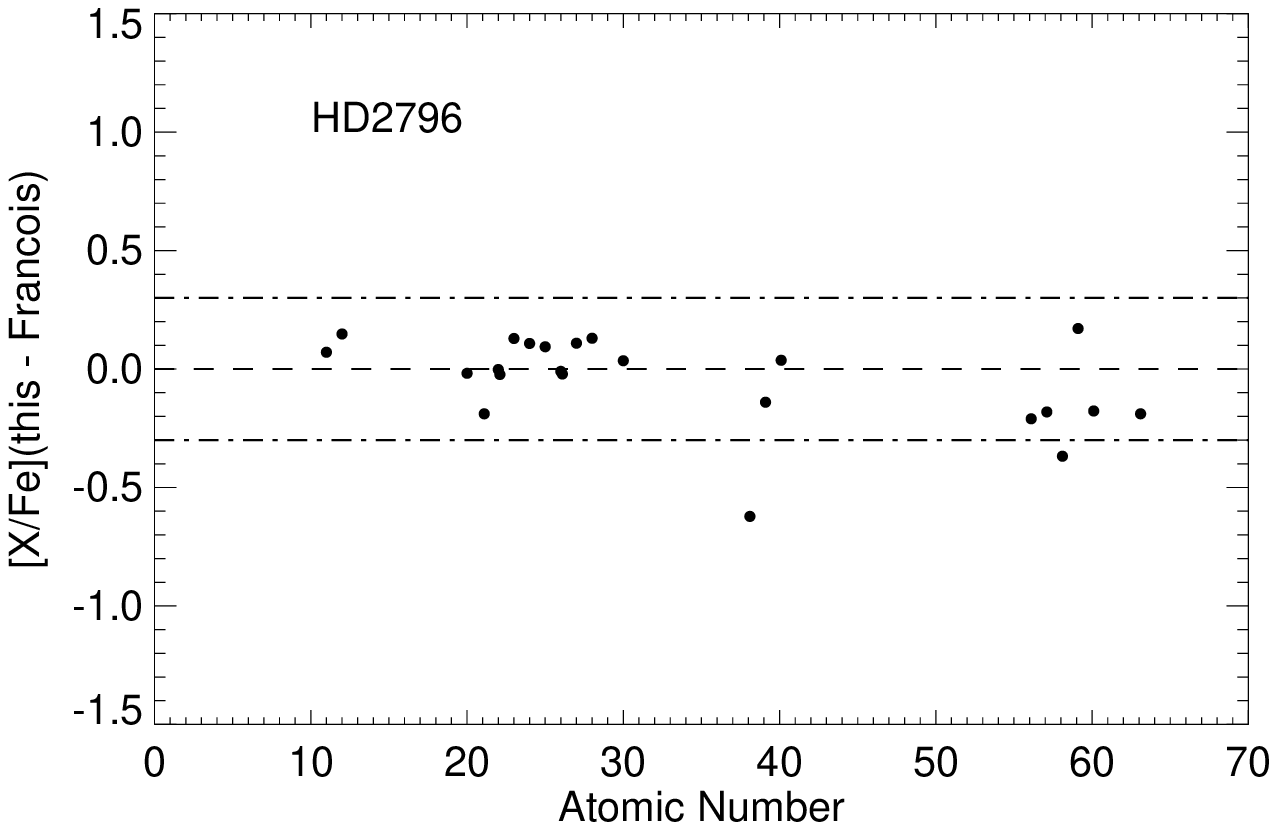}
\caption{Abundances comparison between this work and \citet[for elements with Z$\leq30$]{2004A&A...416.1117C} and \citet[for neutron-capture elements]{2007A&A...476..935F}. 
Excluding Sr abundance, the agreement is highly acceptable. The dashed line refers to $\Delta$= 0 and the dash-dotted lines refer to $\Delta= \pm 0.3 $ dex.}
   \label{fig:Li_difference}
\end{figure}

Table \ref{tab:abun-sigma} lists these abundance uncertainties for HD2796 (as an example), where we varied our atmospheric models effective temperature, 
surface gravity, and  microturbulence by 150K, 0.3dex, and 0.3km s$^{-1}$ respectively. Figure \ref{fig:Li_difference} shows a comparison 
between our elemental abundances and \citet[for elements with Z$\leq30$]{2004A&A...416.1117C} and \citet[for neutron-capture elements]{2007A&A...476..935F}.
Apart from Sr (with $\Delta$ [Sr/Fe]=$-0.62$ relative to Fran{\c c}ois et al.), the differences between our work and their work 
lie within $\pm 0.3$ dex (dash-dotted line in Figure \ref{fig:Li_difference}). We impute our scattering to use cooler temperature,
lower surface gravity, and lower S/N ratio than those used in \citet{2004A&A...416.1117C} and \citet{2007A&A...476..935F} (\Teff = 4950 K, 
 \ensuremath{\log\,g}= 1.5 dex, and S/N ratio = 250-550).

 \section{Results and discussion}\label{sec:discussions}
 
 \subsection{Chemical abundance comparison with literature data}
 
We provided stellar parameters and detailed chemical abundances 
for {five} metal-poor red giant stars, reported for the first time using high resolution spectroscopy. 
These stars exhibit similar chemical abundance patterns to,
reported in other, very and extremely metal-poor stars \citep[e.g.,][]{2007A&A...476..935F, 2013ApJ...762...26Y}.

 \begin{figure}
\epsscale{1.0}
\plotone{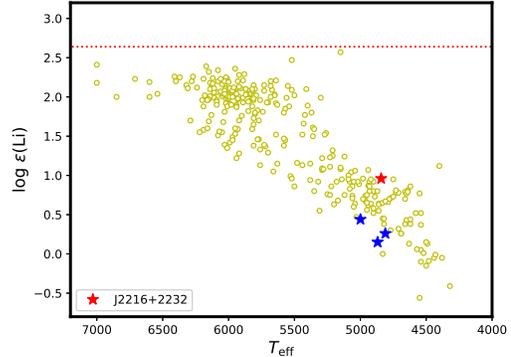}
\caption{log$\epsilon$ lithium abundances as a function of effective temperature. Red filled star refers to J2216$+$2232, {blue filled stars refer to the determined upper limits}, and small open circles 
refer to the full sample of \citet{2014AJ....147..136R}. The dotted line 
shows the predicted primordial lithium abundance, log$\epsilon_{Li}$ = 2.64 \citep{2007ApJS..170..377S}. }
 \label{fig:Li}
\end{figure}

Lithium is considered as a key diagnostic, to test and constrain our understanding of the early Galaxy, of stellar interiors and evolution.
Figure  \ref{fig:Li} illustrates the evolution of lithium as a function of \Teff\,using the halo star sample from \citet[including upper limits]{2014AJ....147..136R}.
The dotted line refers to the primordial lithium abundance predicted by the Standard Big Bang Nucleosynthesis \citep{2007ApJS..170..377S}.
Among the program stars, we could only detect Li in 2216$+$2232 (shown as red filled star) with A(Li) = 0.95. {For completeness the upper limits for 
the rest of our sample stars have been provided (blue filled stars)}. This is not unexpected, as our sample stars are red-giants, whose Li content in the 
outer layers have been diluted by the canonical extra mixing and the first dredge-up (FDU) process.

\begin{figure}
\epsscale{1.2}
\plotone{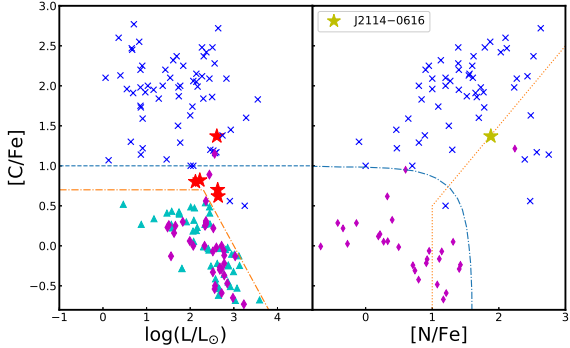}
\caption{Left panel: [C/Fe] versus luminosity for our sample stars. The dash-dotted line indicates the dividing line between 
carbon-enhanced and carbon-normal stars as defined in \citet{2007ApJ...655..492A}. The dashed line corresponds to [C/Fe] = 1.0. 
Right panel: [C/Fe] vs. [N/Fe]. The two criteria for NEMP stars suggested by \citet{2012A&A...547A..76P} are respectively shown in dotted 
([N/Fe] $\geq$ 1.0 and [N/C] $\geq$ 0.5) and dash-dotted lines ([(C+N)/Fe] $>$ 0.9).
The filled red stars refer our sample stars. Non-carbon-enhanced objects studied by previous works \citep{2000A&A...354..169G, 2004A&A...416.1117C, 
2004ApJ...607..474H, 2005ApJ...632..611A} are shown by triangles up. Metal-poor stars from \citet{2005A&A...430..655S} (filled diamonds), 
and CEMP stars from \citet{2007ApJ...655..492A} (crosses) are also plotted for comparison. The candidate to CNEMP objects J2114-0616 is marked.}
\label{fig:C_Fe_L}
\end{figure}

Our sample stars exhibit relatively high [C/Fe] ratios, as shown
in Figure \ref{fig:C_Fe_L} (left panel), which represents [C/Fe] ratio as a function of luminosity. 
We adopted a classification of \citet{2007ApJ...655..492A}, who suggest a
scheme that takes into consideration the nucleosynthesis and
mixing effects in giants. 
We define the stars that satisfy the
following criteria as CEMP stars: [C/Fe]$\geqslant$0.7 for stars with $\log$(L/L$_\odot$)$\leqslant$2.3 and 
[C/Fe]$\geqslant$3.0$-$ $\log$(L/L$_\odot$) for stars with $\log$(L/L$_\odot$)$>$2.3. The luminosities of our stars were calculated 
based on the prescription of \citet{2007ApJ...655..492A}, assuming stellar mass of 0.8 M$_\odot$, following 
\citet{2005ApJ...632..611A} and \citet{2005ApJ...635..349R}. {For completeness, and due to the fact that our sample stars are gaints (see Figure \ref{fig:isoch}), we use the carbon evolutionary correction described in \citet{2014ApJ...797...21P} to assess whether our sample stars could indeed be classified as CEMP, this method suggests that carbon levels decrease as stars evolve into the giant branch phase, due to some level of internal mixing.  As a result, the correction increases the C abundances up to several dex, which support our claims that these stars are CEMP stars.}

With the \citet{2007ApJ...655..492A} definition of CEMP stars and the carbon evolutionary correction described in \citet{2014ApJ...797...21P} in mind, Figure \ref{fig:C_Fe_L} (left panel) shows that our program stars are located above the limit.  
Thus, we point  J1054$+$0528, J1529$+$0804, J1645$+$4357, J2114$-$0616, and J2216$+$2232 as
CEMP stars.

For most of our program stars CN bands are not measurable,
we could only measure N abundance for J2114-0616, which exhibits high nitrogen abundance with [N/Fe]=1.88, [N/C]$>$0.51, and [(C+N)/Fe]=1.53.  
Figure \ref{fig:C_Fe_L} (right panel) shows [C/Fe] as a function of [N/Fe], with the dotted and dash-doted lines referring to \citet{2012A&A...547A..76P} NEMP stars criteria.
We classify J2114$-$0616 as a potential nitrogen-enhanced metal-poor (NEMP) star.
Since J2114$-$0616 satisfies both criteria (star with [C/Fe]$\geqslant$1.0 and [N/C]$\geqslant$0.5), it can be designated as a 
carbon and nitrogen-enhanced metal-poor (CNEMP) star.  

\begin{figure}
\epsscale{1.2}
\plotone{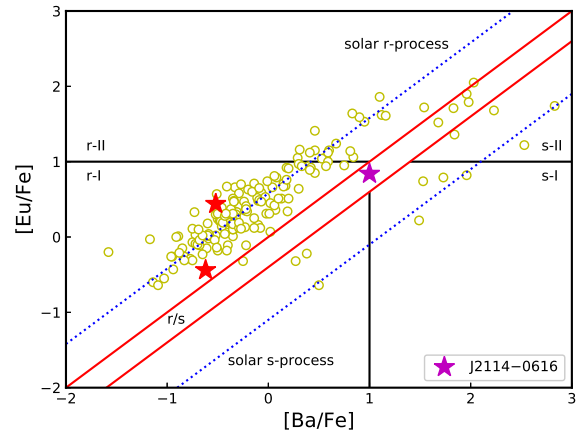}
\caption{[Eu/Fe] as a function of [Ba/Fe], the open circles are data take from saga database \citep{2008PASJ...60.1159S} and the magenta star refer to the position of J2114$-$0616. }.
\label{fig:CEMP}
\end{figure}

Moreover, we investigated [Eu/Fe] as a function of [Ba/Fe] to study
s-process and r-process enrichment, under the pretext that J2114$-$0616 shows 0.0$< $ [Ba/Eu] $ <$ +0.5, and [Ba/Fe] $>$ 0.5 (see figure \ref{fig:CEMP}), we regard it as a CEMP-r/s star.
In addition to the enhancements in both slow (s-) and rapid (r-) process species, J2114$-$0616 shows high [N/Fe] ratio, along with its high [C/Fe],
 which suggest that its peculiar chemical pattern may come from mass transfer from an AGB companion, before it tuned to a white dwarf.

\begin{figure}[t!]
\epsscale{1.1}
\plotone{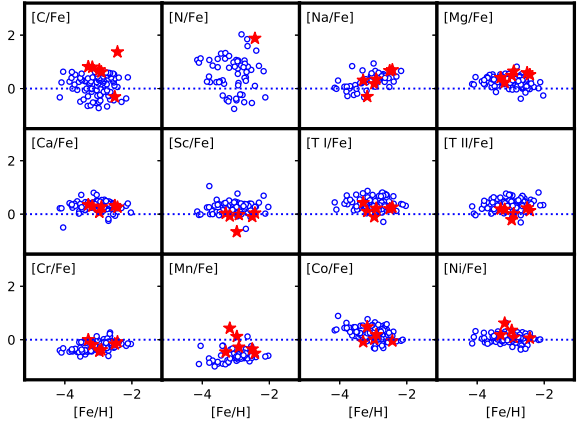}
\caption{{Comparison between the light-element abundances in our sample stars (red filled star) and the literature C-normal metal-poor stars from \citet{2013ApJ...762...26Y}  (blue open circles).}}
\label{fig:light_Fe}
\end{figure}

\begin{figure}
\epsscale{1.2}
\plotone{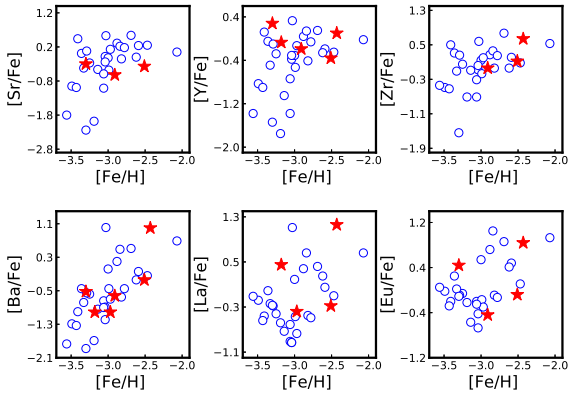}
\caption{{Selected neutron-capture elements abundances in our sample stars (red filled stars), as a function of metallicity, compared to literature data adopted from \citet{2007A&A...476..935F} (blue open circles).}}
\label{fig:heavy_Fe}
\end{figure}

\begin{figure*}[t!]
\epsscale{1.2}
\plotone{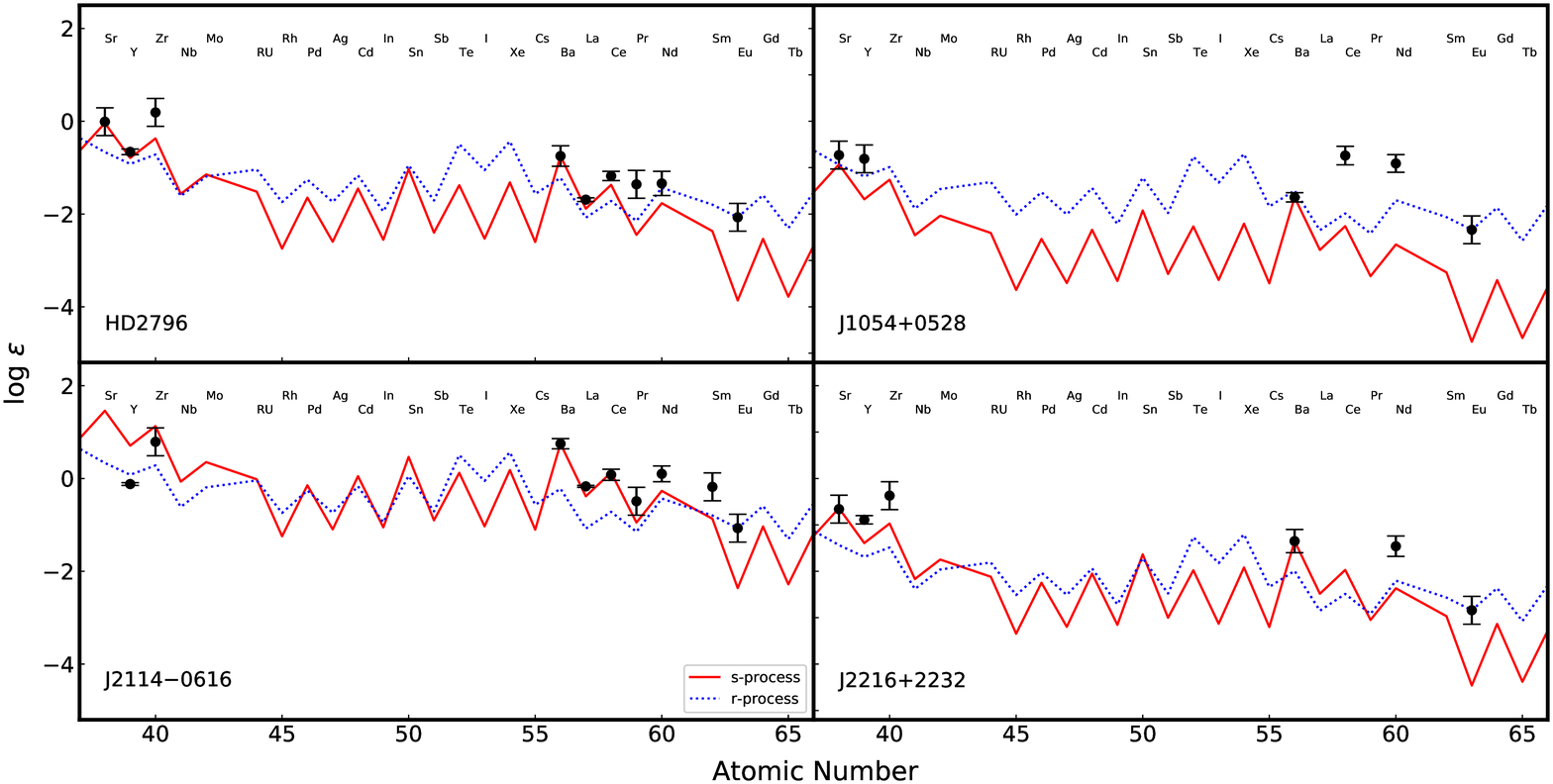}
\caption{Abundance patterns for the neutron-capture elements in our sample. The solid line represents the Solar-system s-process abundance pattern from \citet{2000ApJ...544..302B}, scaled to match the observed abundance of Ba in each star.
The dotted line represents the Solar-system r-process abundance pattern from \citet{2000ApJ...544..302B}, scaled to match the observed abundance of Eu in each star. {Note that we only include these stars with measured Ba and Eu.}}
\label{fig:patterns}
\end{figure*}

{Light element distributions in CEMP stars are quite similar to those in non-carbon-enhanced stars (C-normal). The abundance ratios [X/Fe]
as a function of [Fe/H] of our sample stars (red filled stars) are presented in Figure \ref{fig:light_Fe}, for the elements from C through Zn,
compared with literature data adopted from \citet{2013ApJ...762...26Y} (blue open circles). In general, the abundance ratios seen in our sample show good agreement
with the abundance ratio trends defined by the literature sample. On the other hand, J1529+0804, which shows enhancement
in manganese with [Mn/Fe]=0.43, is not a good example of this agreement, this enhancement may also be true for J1645+4357 ([Mn/Fe]=0.12), keeping in mind that
at low metallicities, the NLTE behavior may systematically increase the manganese abundance up to 0.7 dex (which we will explore in future work) \citep[e.g.,][]{2008A&A...492..823B}}.
  
\subsection{Nucleosynthetic signatures of s- and r-process}

Only elements lighter than zinc can be produced via nuclear-fusions, on the other hand, heavier
elements can be synthesized by either the rapid neutron capture process, r-process, and the slow neutron capture process, s-process
(e.g., \citealt{1994ARA&A..32..153M, 2007PhR...450...97A} and references therein). Metal-poor stars provide unique opportunities to 
attain nucleosynthetic signatures, thus better understanding of the chemical evolution of
these elements and the nucleosynthesis occurred in the early Universe.

As mentioned previously, we were able to determine abundances for up to 10 heavy elements, including the light trans-iron elements ($38\leq$Z$\leq46$) and the second r-process peak elements. 
{The abundances of selected neutron-capture elements for our sample stars (red filled stars), as a function of the metallicity, overlaid with literature data adopted from \citet{2007A&A...476..935F} (blue open circles) are shown in  Figure \ref{fig:heavy_Fe}. No significant discrepancies are found between the selected neutron-capture elements abundances of our sample stars and the literature data}

{Figure \ref{fig:patterns} shows the neutron-capture element abundances for HD2796 and three sample stars, compared with the Solar System s-process (normalized to Ba - solid line) and r-process (normalized to Eu - dotted line) components. The s- and r- fractions were taken from \citet{2000ApJ...544..302B}. Abundances for the first-peak s-process elements (Sr, Y, and Zr) are well described by the Solar s-process for HD2796 and J2216+2232, while for J1054+0528 and J2114-0616 they are roughly consistent with the r-process component. At the same time, the noticeable deviations from for the light elements might interestingly be related to the effects of core-collapse supernovae. In contrast, for the second-peak s-process elements, there is an excellent agreement between measurements and the Solar s-process component for J2114-0616 which, combined with its enhancements in carbon and nitrogen, supports the hypothesis of mass transfer in a binary system from an AGB companion. For J1054+0528, it appears that all the neutron-capture element abundances are a result of an r-process event, with no contributions from the s-process component.}

We could only detect weak \ion{Ba}{2} lines in the spectrum of J1645$+$4357, which results in a relatively low Ba abundance while no other s-process elements can be detected.
Such chemical patterns (high carbon, low barium, and absence of other s-process elements) suggests that J1645$+$4357 was formed out of pristine gas.
The common definition of strongly r-process-enhanced star is [Eu/Fe] $> +1.0$ and [Ba/Eu] $< 0$ and moderately r-process-enhanced star is (r-II stars) $+0.3 \geq$[Eu/Fe]$\leq +1.0$ and [Ba/Eu] $< 0$ \citep[e.g.,][]{2018ARNPS..68..237F}, 
adopting this criterion we suggest that J1054$+$0528 ([Eu/Fe]=0.44) is a new member of the moderately r-process-enhanced stars (r-I stars). On the other hand, J2114$-$0616 exhibits different chemical abundance patterns, 
enhancement of s-process species along with relatively high magnesium abundance
suggesting that $^{22}$Ne($\alpha$, n)$^{25}$Mg may have operate as a main neutron source in J2114$-$0616 \citep{2010A&A...509A..93M}.
Moreover, it turns out that the neutron density linked to this reaction favors the production of cerium (81\% synthesized by s-process) and europium (97\% synthesized by r-process),
suggesting that these elements can't be explained by s-process only, and additional r-process is required to describe this behavior \citep{1998ApJ...497..388G,2000A&A...362..599G}.

\subsection{Kinematics and dynamics} \label{sec:kinematics}

 \begin{figure}
\epsscale{1.2}
\plotone{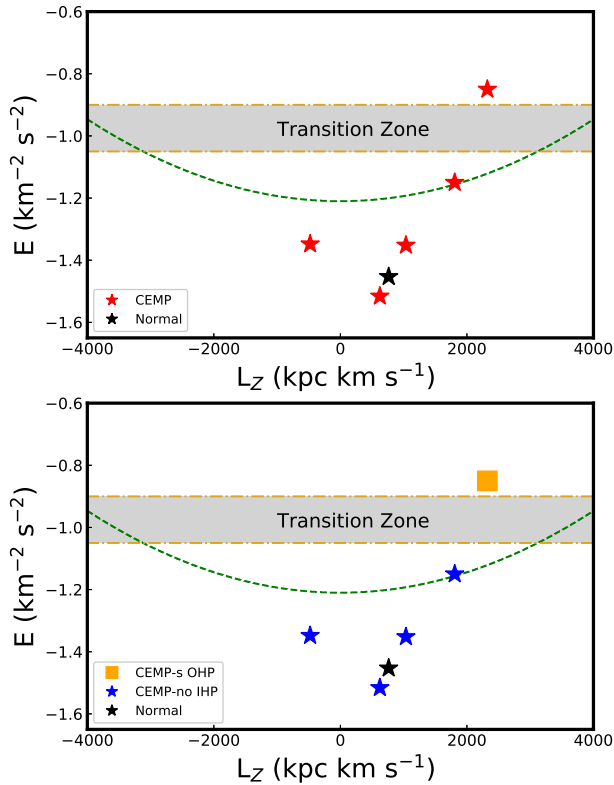}
\caption{Total energy vs. angular momentum in the z direction for the stars in our sample. Top panel: C-normal stars are represented by black
stars, while CEMP stars are denoted by red stars. Bottom panel: the same sample of stars, with blue filled star symbols indicating CEMP-no members of the inner halo, the orange filled square symbol represents the CEMP-no and CEMP-s members of
the outer halo. The green dashed curve denotes
the locus of the points that possess constant apo-Galactic radius, rapo = 15 kpc, while the golden dot-dashed horizontal lines shows the values of the energies
delimiting the transition zone.}
\label{fig:apo_peri}
\end{figure}

\begin{figure}[!ht]
\begin{center}
\includegraphics[scale=0.5]{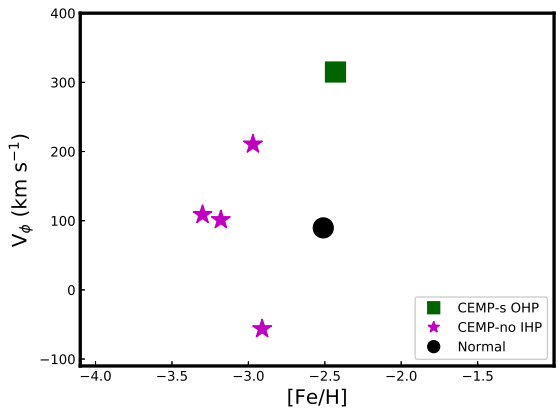}
\end{center}
\caption{Galactocentric rotational velocity vs metallicity for the program stars color coded as in Figure 12.}
\label{fig:vphi}
\end{figure}

The full space motion is derived by combining the observables obtained by Gaia DR2, positions and proper motions ($\alpha$, $\delta$, $\mu_{\alpha}$, $\mu_{\delta}$). We utilize the software {\tt TOPCAT} to cross match our sample with two catalogs from the Virtual Observatory: Gaia DR2 \citep[][for proper motions]{2018A&A...616A...1G} and Gaia DR2 distances \citep[][for distances]{2018AJ....156...58B}, see Table \ref{tab:kinematics_input}. Radial velocities are obtained through cross-correlation with synthetic spectra after the heliocentric corrections to the observed spectra are applied.
The velocities calculated in the Local Standard of Rest (LSR) are referred to as $(U, V, W)$ which are corrected for the motion of the Sun by adopting the values ($U, V, W$) = ($-$9,12,7) km s$^{-1}$ \citep{1981Sci...214..829M}. The velocity component $U$ is taken to be positive in the direction towards the Galactic anti-centre, the $V$ component is positive in the direction towards Galactic rotation, and the $W$ component is positive toward the north Galactic pole. We also compute the rotational velocity component about the Galactic centre in a cylindrical frame, denoted as V$_{\phi}$ , and is calculated assuming  that the LSR is on a circular orbit with a value of 220 km s$^{-1}$ \citep{1986MNRAS.221.1023K}. The orbital parameters are derived by adopting a St\"{a}ckel type gravitational potential (which consists of a flattened, oblate disk, and a nearly spherical massive dark-matter halo; a complete description is given by \citet[][Appendix A]{2000AJ....119.2843C} and integrating their orbital paths based on the starting point obtained from the observations. \\ In addition, we evaluate the integrals of motion for any given orbit, deriving the energy, E, and the angular momentum in the vertical direction, L$_{Z}$ = R x V$_{\phi}$. Note that R represents the distance from the Galactic center projected onto the disk plane. \textbf{Typical errors on the orbital parameters \citep[at Zmax $<$ 50 kpc;][]{2010ApJ...712..692C} are: $\sigma$$_{rperi}$ $\sim$ 1 kpc, $\sigma$$_{rapo}$ $\sim$ 2 kpc, $\sigma$$_{ecc}$ $\sim$ 0.1, $\sigma$$_{Zmax}$ $\sim$ 1 kpc.} \\
\citet{2014ApJ...788..180C} established a method for assigning the membership to the inner- and outer-halo stellar populations based on the integrals of motion (total energy and vertical angular momentum) of a large sample of SDSS/SEGUE DR7 calibration stars. Inner halo stars are mostly highly bound to the Galaxy (lower energy values, $E$ $<$ $-$1.1 km$^{2}$ s$^{-2}$) and possess orbits with apo-galactic distance r$_{\rm apo} <$ 15 kpc, while outer halo stars are less bound to the Galaxy (higher energy values, $E$ $>$ $-$0.9 km$^{2}$ s$^{-2}$ ) and possess orbits with r$_{\rm apo} >$ 15 kpc. Stars with r$_{\rm apo} >$ 15 kpc and $E$ $<$ $-$1.1 km$^{2}$ s$^{-2}$ can be also considered pure inner halo stars. In general, stars in the outer halo are dominated by retrograde orbits but can also possess rotational velocities less retrograde or higlhy prograde, due to the large velocity dispersion  of the outer halo \citep[$\sim$ 165 km s$^{-1}$;][]{2010ApJ...712..692C}. This is clearly evident in the right panel of Figure 4 in \citep{2014ApJ...788..180C} . \\
Figure \ref{fig:apo_peri} shows the total energy, E, as a function of the angular momentum in the vertical direction, L$_{Z}$, for the program stars.  In the top panel, the black filled star symbol represents HD2796, while the red filled star symbols denote the CEMP stars. The grey horizontal area shows the range of binding energy values defining the {\it transition zone} between the inner- and the outer-halo components ($-$1.1 km$^{2}$ s$^{-2}$ $ < E$ $<$ $-$0.9 km$^{2}$ s$^{-2}$), which is defined as the energy range where stars have similar probability to be members of these components. The green dashed curve represents the locus of stars possessing orbits with constant apo-galactic radius r$_{\rm apo}$ = 15 kpc. In the bottom panel the magenta star symbols denote the CEMP-no stars in the inner halo, classified according to their value of binding energy and apo-galactic distance, {the CEMP-r/s star (J2114$-$0616) is represented by an orange filled square and it is member of the outer halo.} \\
Figure \ref{fig:vphi} shows the galactocentric rotational velocity as a function of the metallicity for the program stars. {It is interesting to note that J2114$-$0616 possesses a prograde motion (rotate in the same direction of the galactic disk) with velocities within $\sim$ 2$\sigma$ (CEMP-r/s; J2114$-$0616) of the mean rotational velocity of the outer halo population ($-$80 km s$^{-1}$). Highly prograde stars in the outer halo were also found in the sample of CEMP stars reported in \citet[; Figure 4]{2014ApJ...788..180C}.}\\
Numerical cosmological simulations of MW-mass galaxies predict that stars in the inner halo of the MW formed mainly from massive subgalactic fragments that experienced an extended star formation activity \citep{2009ApJ...702.1058Z, 2011MNRAS.416.2802F, 2012MNRAS.420.2245M, 2013MNRAS.432.3391T, 2014MNRAS.439.3128T}, while outer halo stars formed predominantly in lower-mass subgalactic fragments with short or truncated star formation history 
\citep{2007Natur.450.1020C, 2010ApJ...712..692C, 2012AAS...21922206B,  2013MNRAS.432.3391T, 2014MNRAS.439.3128T, 2016NatPh..12.1170C, 2018ApJ...859L...7C}. The central regions of simulated halos (within $\sim$ 15 kpc) have an important contribution of in-situ stars (formed in the main progenitor galaxy) which have various possible origins \citep{2004ApJ...612..894B, 2009ApJ...702.1058Z,2011MNRAS.416.2802F, 2011MNRAS.415.2652H, 2013MNRAS.432.3391T, 2015MNRAS.454.3185C, 2015ApJ...799..184P, 2016MNRAS.459L..46M}. On the contrary, stars in the outer halo formed primarily in low-mass subgalactic systems which were subsequently accreted. The origin of halo stars can be understood by inspecting a combination of their orbital parameters and integrals of motion. {In case of our sample, J2114$-$0616, possesses orbital parameters, energy and vertical angular momentum that place it in the outer halo population and it likely were formed in low-mass systems outside the virial radius of the progenitor galaxy and accreted later on}. The orbital parameters and binding energy of the remaining CEMP stars, J1054+0528, J1529+0804, J1645+4357 and J2216+2232, suggest that they are members of the inner halo population. However, their metallicity and C-enhancement indicate that they {may have formed} not in situ but in small mass subgalactic fragments which were accreted very early on and contributed to the old central regions of the halo system \citep{2018MNRAS.473.1656T, 2018ApJ...859L...7C}.

 \section{conclusion} \label{sect:conclusion}
In this work we reported on the discovery of {five} CEMP stars selected from the LAMOST DR3 database as metal-poor candidates. 
High-resolution spectra are obtained for the first time with APF /LICK. {We confirmed that J1054$+$0528,
J1529$+$0804,  J1645$+$4357, J2216$+$2232 and J2114$-$0616 show high enhancement in carbon with 
[C/Fe]= 0.82, 0.80, 0.70, 0.62, and 1.37, respectively,} taking an advantage of their observed 
high-resolution spectra and the correction for evolutionary mixing. 
We provided stellar parameters and chemical abundances for up to 25 elements: three light elements (Li, C, N), 
12 elements from Na to Zn, 10 neutron-capture elements (Sr, Y, Zr, Ba, La, Ce, Pr, Nd, Sm, and Eu).  
Our results show no significant abundance differences with literature, thus can be used to study the 
chemical enrichment at the earliest time of the Galaxy.

J1054+0528 shows moderate enhancement in europium [Eu/Fe]=0.44, with low barium-to-iron ratio [Ba/Fe]=$-$0.52. 
These abundances indicate that this star can be classified as CEMP-rI and future higher signal-to-noise observations should be carried out to obtain abundances of other r-process 
elements, such as Gd, Th, Os, Pt, and Pb.

No reliable s-process elements lines were found in J1645$+$4357 spectrum. More accurate measurements of more heavy elements would be needed for further deduction.
Thus, we strongly recommend future observation of J1645$+$4357 in the near-ultraviolet spectroscopy, to determine many key elements such as Ge, Zr, Os, Pt, and Pb.

J2114$-$0616 can be classified as CNEMP-r/s {(carbon nitrogen-enhanced metal-poor)} star, due to the high enhancement in nitrogen with [N/Fe]= 1.88, 
[N/C]= 0.5, barium [Ba/Fe]=1.00, and europium [Eu/Fe]=0.84.  
Other r-process elements, such as Gd, Th, Os, Pt, and Pb can be obtained in future high resolution and high signal-to-noise observations.

Kinematic and dynamic analysis based on Gaia DR2 parameters reveals that {J2114$-$0616 
is} stellar members of the outer halo population and suggests that {it may have formed} in low-mass systems outside 
the virial radius of the progenitor galaxy and accreted later on during the merging and disruption of such systems.
The stars J1054+0528, J1529+0804, J1645+4357 and J2216+2232 are stellar members of the inner halo 
population, however their very low metallicity and enhancement in carbon abundance strongly indicate that they 
were born in low-mass sub-galactic systems which were accreted during the initial phases of Galaxy assembly and 
contributed to the old stellar populations of the inner halo.

\section*{Acknowledgement}

M. K. M. would like to thank The World Academy of Sciences and the Chinese Academy of Sciences for the CAS-TWAS fellowship.
M. K. M. thanks Ian Roederer, Richard de Grijs, and Rene A. Mendez for their comments on earlier versions of the manuscript, 
and the anonymous referee, who made valuable suggestions that helped improve the paper. V. M. P. acknowledges partial support for this work from grant PHY 1430152; Physics Frontier Center/ JINA Center for the Evolution of the Elements (JINA-CEE), awarded by the US National Science Foundation (NSF). S.A. is grateful to China Postdoctoral international exchange program (ISS-SDU) for financial support (Weihai, China).
This study is supported by the National Natural Sciences Foundation of China under grant No. $11890694$, $11573032$, and $11390371$.

\software{MOOG~\citep{1973PhDT.......180S,2011AJ....141..175S}, 
IRAF~\citep{1986SPIE..627..733T, 1993ASPC...52..173T}, NumPy~\citep{2011CSE....13b..22V}, 
SciPy~\citep{SciPy2}, Matplotlib~\citep{Hunter:2007}, Astropy~\citep{2013A&A...558A..33A,2018AJ....156..123A}
DETAIL \citep{detail}, SynthV\_NLTE \citep{2016MNRAS.456.1221R}, binmag \citep{binmag3}}.

\bibliography{APF}
\bibliographystyle{aasjournal}

\begin{deluxetable*}{lllccrrcr}
\tablenum{1}
\tablecolumns{9}
\tabletypesize{\scriptsize}
\tablewidth{0pc}
\tablecaption{Log of the Lick/APF observations.\label{tab:obs}}
\tablehead{
\colhead{}&\colhead{ID}& \colhead{Date}& \colhead{RA}& \colhead{DEC}& \colhead{$r$}& \colhead{Exptime}&
\colhead{S/N}& \colhead{$V_{r}$}\\
\colhead{}& \colhead{}& \colhead{}&  &\colhead{}&\colhead{(mag)}& \colhead{(s)}&
\colhead{(pixel$^{-1})$}& \colhead{(km s$^{-1}$)}}
\startdata
1 &HD2796&    18 Nov 2015 & 00 31 16.91& $-$16 47 40.8&  8.51 &  900*2 &   43& $-$60.51\\
2 &J1054$+$0528& 28 May 2015&10 54 33.10  &$+$05 28 12.7 &12.65 & 1800*4&  42 &82.36\\  
%3 &J1256$+$3440 &30 May 2015 &12 56 42.41& $+$34 40 58.9 &12.64 & 1800*4& 43&284.06\\
3  &J1529$+$0804& 30 May 2015&  15 29 53.94&$+$08 04 48.1&12.49 & 1800*4& 46&23.40\\
4  &J1645$+$4357& 28 May 2015&16 45 14.95 & $+$43 57 12.0 &12.79  & 1800*4& 40&$-$83.20\\
5  &J2114$-$0616 & 23 Sep 2015&21 14 01.52 & $-$06 16 10.3&  10.81 &  1200*4& 46&$-$160.32\\
6 &J2216$+$2232&27 July 2015 &22 16 39.31&  $+$22 32 50.4 & 12.03 & 1800*4& 47&$-$339.38\\
\enddata
\tablecomments{{\footnotesize The S/N ratio per pixel was measured at $\lambda \sim 4500$\,{\AA}\,.\\ 
Luminosity was derived based on \citet{2007ApJ...655..492A} relation}}
\end{deluxetable*}

% stubbed version AAM
\begin{deluxetable*}{lllrrrrrrrrr}
\tablenum{2}
\tablecaption{Equivalent widths of our sample.\label{tab:EWs}}
\tabletypesize{\scriptsize}
\tablecolumns{22}
\tablewidth{0pt}
\tablehead{\colhead{$\lambda$}& \colhead{ Species  }& \colhead{$ \chi$ }& \colhead{\ensuremath{\log\,gf} }&\colhead{HD2796}& \colhead{J1054+0528}&  \colhead{J1529+0804} & \colhead{J1645+4357}   & \colhead{2114$-$0616}         & \colhead{J2216+2232} \\
                                               \colhead{({\AA}\,) }  &      &   \colhead{eV }  & &  \colhead{~~~~m{\AA}\,} &  \colhead{~~~~~~~~~m{\AA}\,}  &  \colhead{~~~~~~~~~m{\AA}\,}  &  \colhead{~~~~~~~~~m{\AA}\,}  &  \colhead{~~~~~~~~~m{\AA}\,}  &  \colhead{~~~~~~~~~m{\AA}\,}   }
\startdata
 4049  &  C(CH)  &     ---  &  ---  &  ---  &  syn &  ---   &     ---  &  syn  &    ---  \\
 4218  &  C(CH)  &     ---  &  ---  & syn &  syn &  ---   &     ---  &  syn  &    syn \\
 4248  &  C(CH)  &     ---  &  ---  &  ---  &  ---  &  syn  &    syn  &  syn  &    ---  \\ 
 4253  &  C(CH)  &     ---  &  ---  & syn &  ---  &  syn  &    ---   &  syn  &    syn \\
 4261  &  C(CH)  &     ---  &  ---  &  ---  &  ---  &   ---  &    syn  &   ---  &    ---  \\
 4273  &  C(CH)  &     ---  &  ---  &   ---  &  syn &   ---  &    syn  &  syn  &    syn \\
 4279  &  C(CH)  &     ---  &  ---  &  ---  &  ---  &   ---  &    syn  &  syn  &    syn \\
 4281  &  C(CH)  &     ---  &  ---  &  syn &  ---  &   ---  &    ---   &  syn  &    syn \\
 4292  &  C(CH)  &     ---  &  ---  &    ---  &  syn &  syn  &    ---   &  syn  &    ---  \\
 4310  &  C(CH)  &     ---  &  ---  &    ---  &  ---  &  syn  &   ---    &   ---  &    syn \\
 4313  &  C(CH)  &     ---  &  ---  &   syn &  ---  &   ---  &    syn  &  syn  &    ---  \\
 4362  &  C(CH)  &     ---  &  ---  &   ---  &  syn &   ---  &    ---   &   ---  &    syn \\
 4366  &  C(CH)  &     ---  &  ---  &    ---  &  ---  &   ---  &    syn  &   ---  &    syn \\
 4214  &  N(CN)  &     ---  &  ---  &    ---  &  ---  &   ---  &    ---   &  syn  &    ---  \\
 6970  &  N(CN)  &     ---  &  ---  &    ---  &  ---  &   ---  &    ---   &  syn  &    ---  \\
5889.95 & \ion{Na}{1} &  0.00 &  0.10 &169.6 &118.7  & 66.4 &174.2 &195.0 &140.0   \\
5895.92 & \ion{Na}{1} &  0.00 & -0.20 &164.8 &120.4  & 65.1 &161.1 &181.8 &159.4   \\
4702.99 & \ion{Mg}{1} &  4.33 & -0.44 & 77.2 & 26.1     &  8.8 & 64.5 & 73.0 & 51.3   \\
5172.68 & \ion{Mg}{1} &  2.71 & -0.45 & 217.9 &124.4  &209.7 &211.0&183.6   \\
5183.60 & \ion{Mg}{1} &  2.72 & -0.24 &260.2 &145.6&147.8 &232.1 &267.0 &218.5   \\
5528.40 & \ion{Mg}{1} &  4.35 & -0.50 & 68.7 & 36.2  & 33.7 & 72.7 & 81.3 & 48.0   \\
5711.09 & \ion{Mg}{1} &  4.35 & -1.72 &  3.5 & 10.5  &--- &--- &  9.7 &  5.3   \\
4283.01 &  \ion{Ca}{1} &  1.89 & -0.22 & 60.9 &---  & 15.2 & 54.9 & 78.5 & 59.9   \\
4318.65 &  \ion{Ca}{1} &  1.89 & -0.21 & 64.7 &---  &---  &---  & 22.2 & 40.0   \\
4425.44 &  \ion{Ca}{1} &  1.88 & -0.36 & 50.9 & 27.2 & 19.3 &---  & 52.9 &---    \\
4454.78 &  \ion{Ca}{1} &  1.90 &  0.26 & 81.9 & 31.9 & 46.5 & 42.1 & 76.4 & 64.9   \\
4455.89 &  \ion{Ca}{1} &  1.90 & -0.53 & 43.9 &---  & 18.8 &---  & 38.7 & 15.9   \\
\enddata
\tablecomments{Table \ref{tab:EWs} is published in its entirety in the machine-readable format. A portion is shown here for guidance regarding its form and content.}
\end{deluxetable*}

\begin{deluxetable*}{lllllllllllllllllllllll}
\tablenum{3}
\tablecolumns{22}
\tabletypesize{\scriptsize}
\tablewidth{0pc}
\tablecaption{Stellar Parameters of the Program Stars.\label{tab:stellar-param}}
\tablehead{
\colhead{}&\multicolumn{4}{c}{Lick/APF (adopted)}&\multicolumn{4}{c}{LAMOST}&&\multicolumn{4}{c}{Gaia DR2}&&\multicolumn{2}{l}{Photometry }&&&\multicolumn{2}{l}{Luminosity}&&\\
\cline{2-5}\cline{7-9}\cline{12-13}\cline{16-17}\cline{20-21}\\
\colhead{ID}& \colhead{\Teff }& \colhead{\ensuremath{\log\,g}}& \colhead{[Fe/H]}& \colhead{$\xi$} &
\colhead{}& \colhead{\Teff}& \colhead{\ensuremath{\log\,g}}& \colhead{[Fe/H]}& & &\colhead{\Teff}& \colhead{\ensuremath{\log\,g}}& && &\colhead{\Teff}&  &&\colhead{$~~L_{\ast}/ L_{\odot}$} \\
\colhead{}& \colhead{(K)}& \colhead{(cgs)}& \colhead{}& \colhead{(kms$^{-1}$)} &
\colhead{}& \colhead{(K)}& \colhead{(cgs)}& &\colhead{}&\colhead{}& \colhead{(K)}& \colhead{(cgs)}&&&& \colhead{(K)} &&& \colhead{{~~(ergs$^{-1}$)}}}
\startdata
HD2796           & 4869       &      1.04   &     $-$2.51&  2.01&&		...&...&...&&&				4995&2.15 &&&&~~~~...         &&&~~~~1014.0\\
J1054$+$0528&  5030	&	1.88	  &	$-$3.30&	1.94&&		5094& 1.76& $-$3.23&&&		5080&1.89 &&&&~~~~4981    &&&~~~~166.94\\ 
%J1256$+$3440&  5215	&	1.74   &	$-$2.76&	2.37&&		5238 &1.75& $-$3.02&&&		5298&1.90 &&&&~~~~5004    &&&~~~~266.26 \\
J1529$+$0804 &  5085	&	2.00   &   $-$3.18&	2.34&&		5026 &1.54 &$-$3.27&&&		5041&1.90 &&&&~~~~4913    &&&~~~~132.27\\ 
J1645$+$4357 & 4810      &	1.39   &	$-$2.97&	2.93&&		4715 &2.14 &$-$3.05&&&		4886&1.39 &&&&~~~~4652    &&&~~~~431.38\\
J2114$-$0616 &  4999	&      1.48	  &	$-$2.43&	2.11&&		4377 & 1.55& $-$2.95&&&	4870&1.66 &&&&~~~~4831    &&&~~~~409.09\\
J2216$+$2232 &  4842     &	1.40	  &	$-$2.91&	1.79&&		4902 &2.65 &$-$3.13&&&		5018&2.05 &&&&~~~~4815    &&&~~~~432.89  \\
\enddata
\tablecomments{{The tenth column was measured using parallaxes adopted from Gaia DR2 and stellar mass of $\sim$ 0.8 M$_{\odot}$}}
\end{deluxetable*}

\begin{deluxetable*}{lrrcrcrrcrcrrcrcrrcrcc}
\tablecolumns{22}
\tablenum{4}
\tabletypesize{\scriptsize}
\tablewidth{0pt}
\tablecaption{LTE Abundances of Individual Elements for the Program Stars. \label{tab:res}}
\tablehead{
\colhead{}& \multicolumn{4}{c}{HD2796}& \colhead{}&
            \multicolumn{4}{c}{J1054$+$0528}& \colhead{}&
            \multicolumn{4}{c}{J1529$+$0804}& \colhead{}\\
\cline{2-5}\cline{7-10}\cline{12-15}\\
\colhead{}& \colhead{log\,$\epsilon$(X)}& \colhead{[X/Fe]}& \colhead{$\sigma$}& \colhead{$N$}& \colhead{}&
            \colhead{log\,$\epsilon$(X)}& \colhead{[X/Fe]}& \colhead{$\sigma$}& \colhead{$N$}& \colhead{}&
	    \colhead{log\,$\epsilon$(X)}& \colhead{[X/Fe]}& \colhead{$\sigma$}& \colhead{$N$}& \colhead{}&}
\startdata
\ion{Li}{1}& ...&...&...&...&&...&...&...&...&&...&...&...&... \\
C(CH)             & 5.60&$-$0.31&0.05&5&&5.95&0.82&0.09&6&&6.05&0.80&0.11&4\\
~~(CH)$_{corr}$              &...&0.74&...&...&&  ...&0.04&...&...&&...&0.01&...&... \\
N(CN)             &...&...&...&...&&  ...&...&...&...&&...&...&...&... \\
\ion{Na}{1}&     4.40& 0.67 &0.05&2     &&3.25&0.31 &0.05&2       &&  2.76&$-$0.30& 0.24&2 \\
\ion{Mg}{1}&      5.68 & 0.59&0.05 &4       &&    4.71 &0.41&0.12 &5&&     4.64 &0.22&0.10 &4\\
\ion{Ca}{1}&     4.15     &  0.32     &  0.10     &  21&&3.39     &  0.35     &  0.29     &  11& & 3.46   &   0.30    &  0.18 &       17\\
\ion{Sc}{2}&       0.55     & $-$0.09     &  0.05     &  13&& $-$0.12     &  0.03     &  0.24     &   6&& $-$0.10 &   $-$0.07&      0.05&        3\\
\ion{Ti}{1}&          2.64     &  0.20     &  0.09     &  18&&  2.08     &  0.43     &  0.25     &  13&&  1.87 &      0.10&     0.30 &      11\\
\ion{Ti}{2}&           2.67     &  0.23     &  0.12     &  33&& 1.84     &  0.19     &  0.30     &  25&&  1.95  &    0.18  &    0.32   &    24\\
\ion{V}{1}&         1.36     & $-$0.06     &  0.11     &   1&&   0.93     &  0.30     &  0.13     &   1&&  ...&...&...&...\\
\ion{Cr}{1}&         2.97     & $-$0.16     &  0.10     &   9&& 2.34     &  0.00     &  0.30     &   5&& 2.27 &     $-$0.19&     0.29&        8\\
\ion{Mn}{1}&         2.57     & $-$0.35     &  0.12     &   3&& 1.67     &  $-$0.46     &  0.10    &   1&& 2.68&      0.43&     0.02&        2\\
\ion{Fe}{1}&         4.99     & 0.00     &  0.12     & 217&& 4.20     &  0.00     &  0.16     &  84&& 4.32&     0.00&      0.26&      146\\
\ion{Fe}{2}&        4.99     & 0.00     &  0.09     &  24 &&4.20     &  0.00     &  0.18     &   8 &&  4.32&     0.00&      0.25&       17\\
\ion{Co}{1}&        ...&...&...&...&&1.61     & $-$0.08     &  0.20     &   2&&2.29&      0.48&      0.06&        3\\
\ion{Ni}{1}&        ...&...&...&...&&  3.10     &  0.18     &  0.13     &   4&& 3.66&      0.62&      0.32&        7\\
\ion{Zn}{1}&         2.33     &  0.28     &  0.03     &   2&&...&...&...&...&&1.95&      0.57&      0.12&       1\\
\ion{Sr}{2}&      $-$0.01     & $-$0.37     &  0.10     &   1&& $-$0.73     & $-$0.30     &  0.13     &   1&&...&...&...&...\\
\ion{Y}{2}&      $-$0.66     & $-$0.36     &  0.06     &   3&& $-$0.81     &  0.28     &  0.16     &   1&& $-$1.04     &$-$0.07&      0.19&        2\\
\ion{Zr}{2}&      0.19     &  0.12     &  0.12     &   1&&...&...&...&...&&...&...&...&...\\
\ion{Ba}{2}&       $-$0.57     & $-$0.24     &  0.22     &   3&&  $-$1.64     & $-$0.52     &  0.10     &   3&&$-$2.01&     $-$1.01&      0.13&        2\\
\ion{La}{2}&      $-$1.69     & $-$0.28     &  0.04     &   2&&...&...&...&...&&$-$1.63&      0.45&      0.13&        1\\
\ion{Ce}{2}&       $-$1.18     & $-$0.25     &  0.10     &   3&& $-$0.74     &  0.98     &  0.20     &   2&&$-$0.77&      0.84&      0.11&       2\\
\ion{Pr}{2}&        $-$1.36     &  0.43     &  0.17     &   1&&...&...&...&...&&$-$0.88&      1.58&      0.19&        1\\
\ion{Nd}{2}&          $-$1.34     & $-$0.25     &  0.26     &   7&& $-$0.91     &  0.97     &  0.19     &   9&& $-$0.71&      1.05&     0.58&       5\\
\ion{Sm}{2}&  ...&...&...&...&&...&...&...&...&&$-$0.60&      1.62&     0.05&        2 \\
\ion{Eu}{2}&  $-$2.07     & $-$0.08     &  0.11    &   1&&$-$2.34     &  0.44     &  0.14     &   1&& ...&...&...&...\\
\tableline
\enddata
\clearpage
\end{deluxetable*}

\begin{deluxetable*}{lrrcrcrrcrcrrcrcrrcrcc}
\tablecolumns{22}
\tablenum{4}
\tabletypesize{\scriptsize}
\tablewidth{0pt}
\tablecaption{Continued.}
\tablehead{
\colhead{}& \multicolumn{4}{c}{J1645$+$4357}& \colhead{}&
            \multicolumn{4}{c}{J2114$-$0616}& \colhead{}&
            \multicolumn{4}{c}{J2216$+$2232}& \colhead{}&
            \multicolumn{4}{c}{Sun}& \colhead{}\\
\cline{2-5}\cline{7-10}\cline{12-15}\cline{17-20}\\
\colhead{}& \colhead{log\,$\epsilon$(X)}& \colhead{[X/Fe]}& \colhead{$\sigma$}& \colhead{$N$}& \colhead{}&
            \colhead{log\,$\epsilon$(X)}& \colhead{[X/Fe]}& \colhead{$\sigma$}& \colhead{$N$}& \colhead{}&
	    \colhead{log\,$\epsilon$(X)}& \colhead{[X/Fe]}& \colhead{$\sigma$}& \colhead{$N$}& \colhead{}&
	    \colhead{}& \colhead{log\,$\epsilon$(X)}& \colhead{}& \colhead{}& \colhead{}&}
\startdata
\ion{Li}{1} &...&...&...&...&&...&...&...&... &&0.96&2.82&0.04&1&&& 1.05\\
C(CH)             &  6.16&0.70&0.04&6&& 7.37&1.37& 0.05& 10&& 6.14&0.62& 0.05& 8  &&  &8.43  \\
~~(CH)$_{corr}$             &...&0.45&...&...&&  ...&0.17&...&...&&...&0.45&...&... &  &&...\\
N(CN)             &...&...&...&...&&  7.28&1.88&0.17&2&& ...&...&...&... &&  &7.83\\
\ion{Na}{1}&3.45&0.18&0.04&2       &&4.69&0.88 &0.05&2     &&3.68&0.35 &0.05&2     &  &&6.24 \\
\ion{Mg}{1}&      5.12 & 0.49&0.05 &4&&    5.69 &0.52&0.12 &5&&    5.34 &0.65&0.02& 4&&& 7.60  \\
\ion{Ca}{1}&     3.44     &  0.07     &  0.28     &  11&&4.17     &  0.26     &  0.29     &  20&&3.70     & 0.27     &  0.17     &  18     && &6.34  \\
\ion{Sc}{2}&   $-$0.48    & $-$0.66     &  0.23     &   4&& 0.75     &  0.04     &  0.25     &  13&& 0.22     & $-$0.02     &  0.14     &   8    && &3.15   \\
\ion{Ti}{1}&     1.88     & $-$0.10     &  0.26     &  11&& 2.78     &  0.26     &  0.16     &  15&& 2.26     & 0.22      &  0.13     &  15       && &4.95 \\
\ion{Ti}{2}&         1.76     & $-$0.21     &  0.25     &  25&&  2.65     &  0.13     &  0.31     &  30&& 2.05     &   0.01     &  0.27     &  26    &&  & 4.95 \\
\ion{V}{1}&         ...&...&...&...&& ...&...&...&...&& ...&...&...&...&&& 3.93 \\
\ion{Cr}{1}&          2.24     & $-$0.43     &  0.30     &   6 && 3.13     & $-$0.08     &  0.09     &   8&& 2.35     & $-$0.38     &  0.22     &  10     && &5.64  \\
\ion{Mn}{1}&         2.58     &  0.12     &  0.08     &   2&& 2.49     & $-$0.51     &  0.03     &   2&& 2.24     & $-$0.28     &  0.21     &   2      &&&  5.43 \\
\ion{Fe}{1}&       4.53     & 0.00     &  0.22     & 150&& 5.07     & 0.00     &  0.14     & 156&& 4.58     & 0.00     &  0.18     & 181    &&& 7.50   \\
\ion{Fe}{2}&        4.53     &0.00     &  0.20     &  18&&  5.07     & 0.00     &  0.17     &  24&&  4.58     & 0.00     &  0.19     &  20  &&& 7.50     \\
\ion{Co}{1}&       2.04     &  0.02     &  0.09     &   1&& 2.51     & $-$0.05     &  0.03     &   2&& 2.27     & 0.19    &  0.11     &   3     &&& 4.99  \\
\ion{Ni}{1}&       3.60     &  0.35     &  0.16     &   8&& 3.85     &  0.06     &  0.10     &  11&& 3.40     & 0.09     &  0.15     &   9  &&&  6.22   \\
\ion{Zn}{1}&    ...&...&...&...&& 2.43     &  0.30     &  0.12     &   2&& 1.92     & 0.27     &  0.04     &   2  &&&4.56     \\    
\ion{Sr}{2}&       ...&...&...&...&&  ...&...&...&...&& $-$0.66     & $-$0.62    &  0.08     &   1   &&& 2.87    \\
\ion{Y}{2}&    ...&...&...&...&& $-$0.12     &  0.10     &  0.03     &   2&&$-$0.89     & $-$0.19    &  0.09     &   2  &&&2.21     \\ 
\ion{Zr}{2}&    ...&...&...&...&& 0.79     &  0.64     &  0.13     &   1&& $-$0.37     & $-$0.04     &  0.10     &   1     &&&2.58   \\
\ion{Ba}{2}&     $-$1.80     & $-$1.01     &  0.15     &   3&&  0.75     &  1.00     &  0.11     &   3&& $-$1.35     & $-$0.62      &  0.25     &   3   &&& 2.18    \\
\ion{La}{2}&      $-$2.25     & $-$0.38     &  0.11     &   1&& $-$0.17     &  1.16     &  0.02     &   2 &&...&...&...&...&& &1.10\\
\ion{Ce}{2}&      ...&...&...&...&&0.08     &  0.93     &  0.12     &   3&&...&...&...&...&&&  1.58\\
\ion{Pr}{2}&   ...&...&...&...&& $-$0.49     &  1.22     &  0.15    &1 &&   ...&...&...&...   && &0.72 \\
\ion{Nd}{2}&  ...&...&...&...&&    0.10     &  1.11     &  0.17     &   8&&$-$1.46     & 0.03    &  0.22     &   5    &&&1.42   \\
\ion{Sm}{2}& ...&...&...&...&& $-$0.18     &  1.29     &  0.12     &   1&&...&...&...&...&&& 0.96\\
\ion{Eu}{2}&    ...&...&...&...&& $-$1.07     &  0.84     &  0.13     &   1&&$-$2.84     & $-$0.44     &  0.12    &   1    &&&1.07   \\
\tableline
\enddata
\clearpage
\tablecomments{{N refers to the number of lines adopted for determination of the elemental abundances.}}
\end{deluxetable*}

\begin{deluxetable*}{lccccrrrrccccr}
\tablenum{5}
\tabletypesize{\scriptsize}
\tablewidth{0pt}
\tablecaption{Li, Na, Mg NLTE abundances}\label{tab:nlte}
\tablehead{
\colhead{Star}& \multicolumn2c{log$\epsilon$(Li)} & \multicolumn2c{log$\epsilon$(Na)} & \colhead{[Na/Fe]$_{NLTE}$ } &
\multicolumn2c{log$\epsilon$(Mg)} & \colhead{[Mg/Fe]$_{NLTE}$}
\\
\colhead{} & \colhead{LTE} & \colhead{NLTE} & \colhead{LTE} & \colhead{NLTE} & \colhead{} & \colhead{LTE} & \colhead{NLTE} & \colhead{} 
}
\startdata
HD2796   &          $<$ 0.14        &     $<$ 0.15               &   4.40   &  3.65                  & $-$0.08    & 5.68 & 5.69 & 0.60  \\
J1054$+$0528 & ~~~...       & ~~~...                   &   3.25   &  2.91                  & $-$0.03    & 4.71 & 4.81 & 0.51  \\
%J1256$+$3440  & $<$ 0.50       &  $<$ 0.52                  &   3.64   &  3.10                  &  $-$0.35   & 5.09 & 5.13 & 0.32  \\
J1529$+$0804&$<$ 0.30         &   $<$ 0.32                 &   2.76   &  2.53                  & $-$0.53    & 4.64 & 4.75 & 0.33  \\
J1645$+$4357&$<$ 0.24        &  $<$ 0.26                  &   3.45   &  2.97                  & $-$0.30    & 5.12 & 5.10 & 0.47  \\
J2114$-$0616&$<$ 0.44       &     $<$ 0.44               &   4.69   &  3.88                  &  0.07    & 5.69 & 5.69 & 0.52   \\ 
J2216$+$2232&      ~~~0.95 & ~~~0.96                                  &   3.68   &  3.06                  & $-$0.27    & 5.34 & 5.33 & 0.64  \\ 
 \tableline
\enddata
\tablecomments{{J1054$+$0528 has defect in the spectrum at the region of the \ion{Li}{1} line.}}
\end{deluxetable*}

\begin{deluxetable}{lrrr}
\tablenum{6}
\tablecolumns{4}
\tabletypesize{\scriptsize}
\tablewidth{0pc}
\tablecaption{Uncertainties of log\,$\epsilon$(X) Propagated from the Stellar Parameters
(as Described in Section~\ref{subsec:abuns_uncer}),
Computed for HD2796 as an Example.\label{tab:abun-sigma}}
\tablehead{
\colhead{Ion}& \colhead{$\Delta$Teff}& \colhead{$\Delta$\ensuremath{\log\,g}}& \colhead{$\Delta \xi$}\\
\colhead{}& \colhead{$+$ 150\,K}& \colhead{$+$ 0.3\,dex}& \colhead{$+$ 0.3\,km s$^{-1}$}}
\startdata
CH(C)       &0.32    &       $-$0.12&       0.02\\
CN(N)       &0.50          & $-$0.10      & 0.01\\
\ion{Na}{1} &0.16 &          $-$0.05   &     $-$0.03\\
\ion{Mg}{1} &0.14          &$-$0.11        &$-$0.03 \\
\ion{Ca}{1} & 0.05 &     $-$0.02 &        $-$0.11\\
\ion{Sc}{2} &0.09 &       $-$0.12 &        $-$0.13\\
\ion{Ti}{1} & $-$0.03 &     0.00 &       $-$0.02\\
\ion{Ti}{2} & 0.11 &       -0.12&       -0.13\\
\ion{V}{2}  &$-$0.04 &      0.00 &     $-$0.02\\
\ion{Cr}{2} & $-$0.02 &        $-$0.01 &        $-$0.01\\
\ion{Mn}{1} &0.01 &       $-$0.01 &      $-$0.08\\
\ion{Fe}{1} & 0.16 &    $-$0.03 &   $-$0.15 \\
\ion{Fe}{2} &0.15 &       $-$0.13&       $-$0.219\\
\ion{Co}{1} &   $-$0.04 &        0.01 &      0.03\\
\ion{Ni}{1} &$-$0.01 &      $-$0.05 &       $-$0.13\\
\ion{Zn}{1} &0.06 &        $-$0.06 &     $-$0.18\\
\ion{Sr}{2} &0.06 &      $-$0.07 &      0.08\\
\ion{Y}{2}  &0.07 &       $-$0.13 &      $-$0.17\\
\ion{Zr}{1}&  0.07 &       $ -$0.12 &      $-$0.16\\
\ion{Ba}{2} &0.02 &        $-$0.12 &       0.08\\
\ion{La}{2}& 0.04 &      $-$0.12&       $-$0.14\\
\ion{Ce}{2}&0.05 &      $-$0.12 &       $-$0.15\\
\ion{Pr}{2}& 0.03 &       $-$0.12 &      $-$0.13\\
\ion{Nd}{2}&0.03 &     $-$0.12 &      $-$0.13\\
\ion{Eu}{2}&0.05 &       $-$0.12 &        $-$0.14\\
\enddata
\end{deluxetable}

\begin{deluxetable*}{lcccccccccccccccccr}
\tablenum{7}
\tablecaption{Parallaxes, Proper Motions and Distances  \label{tab:kinematics_input}}
\tablewidth{0pc}
\tablehead{
\colhead{Star} &\colhead{\textit{Gaia} DR2 source ID} &\colhead{$\varpi$} &\colhead{error} & &\colhead{pmra} &
\colhead{error} &  &\colhead{pmdec} &\colhead{error} &  &\colhead{Distance} &\colhead{d1} &\colhead{d2} & &
\cline{3-4}\cline{6-7}\cline{9-10}\cline{12-14}\cline{16-17}
 &\colhead{} &\multicolumn{2}{c}{(mas)} &  &\multicolumn{2}{c}{(mas~yr$^{-1}$)} &  &
\multicolumn{2}{c}{(mas~yr$^{-1}$)} &  &\multicolumn{3}{c}{(kpc)}}
\startdata
HD2796                     & 2367454697327877504 & 1.4859 & 0.0626   & &     $-$1.375  & 0.164 & &$-$51.052 & 0.084 & & 0.661 & 0.027 & 0.028    \\        
J1054$+$0528          & 3864140775805950208 & 0.2149 & 0.0393   & &     $-$8.678  & 0.073 & &$-$  4.183 & 0.055 & & 3.548 & 0.404 & 0.502     \\
%J1256$+$3440          & 1515900293283319168 & 0.1945 & 0.0469   & &           4.068 & 0.058 & &$-$13.785& 0.048 & & 3.601 & 0.467 & 0.593     \\
J1529$+$0804          & 1164484488577137792 & 0.2365 & 0.0362   & &      $-$2.401 & 0.056 & &$-$9.169  & 0.05   & & 3.628 & 0.425 & 0.544     \\
J1645$+$4357          & 1357725650023190784 & 0.0454 & 0.0207   & &           0.327 & 0.033 & &$-$5.663  & 0.042 & & 8.355 & 0.996 & 1.215     \\
J2114$-$0616           & 6910940758263238912 & 0.4496 & 0.0373   & &    $-$20.444 & 0.067 & &$-$6.376  & 0.068 & & 2.082 & 0.153 & 0.178     \\
J2216$+$2232          & 1878089211702170880 & 0.3721 & 0.0384   & &      $-$4.454 & 0.056 & &$-$1.477 & 0.058  & & 2.440 & 0.211 & 0.253     \\
\enddata
\tablecomments{The d1 and d2 columns indicate the 16th percentile and 84th percentile confidence intervals.}
\end{deluxetable*}

\begin{deluxetable*}{lrrrrrrrrrr}
\tablenum{8}
\tablecolumns{10}
\tabletypesize{\scriptsize}
\tablewidth{0pc}
\tablecaption{Kinematics parameters \label{tab:kinematics}}
\tablehead{
\colhead{Star ID}& \colhead{V$_{R}$}& \colhead{V$_{\Phi}$}&  \colhead{V$_{Z}$}& \colhead{V$_{\perp}$}& \colhead{Z$_{max}$}& \colhead{R$_{apo}$} & \colhead{R$_{peri}$} & \colhead{e}
& \colhead{E} & \colhead{L$_{z}$}  }
\startdata	
HD2796            &   -95.03&     89.62&     37.81&    102.28&          1.02&      9.59&      2.28&      0.62&     -1.45&    763.40      \\
J1054$+$0528 &   82.98&    108.74&    -15.88&     84.49&          3.59&     10.81&      3.60&      0.50&     -1.35&   1036.00      \\
%J1256$+$3440 &   -174.71&    119.41&    318.02&    362.85&         38.40&     40.25&      9.10&      0.63&     -0.73&   1040.00      \\
J1529$+$0804 &  -86.75&    101.22&     -8.18&     87.13&          3.01&      7.43&      2.52&      0.50&     -1.52&    624.00      \\
J1645$+$4357 &  -74.40&    210.51&    -58.18&     94.45&          7.67&     14.50&      7.82&      0.30&     -1.15&   1807.00      \\
J2114$-$0616  & -220.45&    314.95&     37.41&    223.60&          4.70&     33.16&      5.64&      0.71&     -0.85&   2323.00      \\
J2216$+$2232 &  -35.03&    -56.42&    182.30&    185.64&          8.70&      9.22&      4.08&      0.41&     -1.35&   -478.40      \\
\enddata
\end{deluxetable*}

\end{document}